\documentclass[10pt,reprint,prl,floatfix,superscriptaddress]{revtex4-2}  

\usepackage{graphicx}
\usepackage{siunitx}
\usepackage{textcomp}

\usepackage{gensymb}
\usepackage{amsmath,amssymb}
\usepackage{upgreek}
\usepackage{bm}
\usepackage{mathrsfs}
\usepackage{color}
\usepackage[normalem]{ulem}
\usepackage{soul} 
\usepackage{hyperref}
\usepackage{pslatex}

\definecolor{mymagenta}{rgb}{0.6,0,0.6}
\definecolor{myblue}{rgb}{0,0,0.7}

\begin{document}  
\title{Metasurface lasers programmed by optical pump patterns}

\author{Nelson de Gaay Fortman}
    \affiliation{Institute of Physics, University of Amsterdam, 1098 XH Amsterdam, The Netherlands} 
    \affiliation{Department of Physics of Information in Matter and Center for Nanophotonics, NWO-I Institute AMOLF, Science Park 104, NL1098XG Amsterdam, The Netherlands}
   \author{Radoslaw Kolkowski}
    \affiliation{Department of Applied Physics, Aalto University, P.O. Box 13500, FI-00076 Aalto, Finland}
\author{Nick Feldman}
    \affiliation{Department of Physics of Information in Matter and Center for Nanophotonics, NWO-I Institute AMOLF, Science Park 104, NL1098XG Amsterdam, The Netherlands}
    \affiliation{Advanced Research Center for Nanolithography (ARCNL), Science Park 106, 1098 XG Amsterdam, The Netherlands}
\author{Peter Schall}
    \affiliation{Institute of Physics, University of Amsterdam, 1098 XH Amsterdam, The Netherlands}
\author{A. Femius Koenderink\footnote[1]{Corresponding author: f.koenderink@amolf.nl} }
    \affiliation{Department of Physics of Information in Matter and Center for Nanophotonics, NWO-I Institute AMOLF, Science Park 104, NL1098XG Amsterdam, The Netherlands}
    \affiliation{Institute of Physics, University of Amsterdam, 1098 XH Amsterdam, The Netherlands}

\date{\today}

\begin{abstract}
Metasurface lasers offer unprecedented control over light emission, yet their spatial and modal characteristics are typically fixed post-fabrication. Here, we introduce a reconfigurable plasmonic metasurface laser platform in which the lasing area geometry, and thus the emission properties, are dynamically programmed via spatially structured optical pumping. Using hexagonal arrays of silver nanoparticles embedded in dye-doped waveguides, we demonstrate lasing at high-symmetry points of the Brillouin zone, including the $\mathit{K}$ and $\mathit{M}$ points. $\mathit{K}$-point lasing exhibits spontaneous symmetry breaking (SSB) in relative intensity between degenerate $\mathit{K}$ and $\mathit{K'}$ modes, with no bias induced by pump geometry, even for geometries that explicitly break symmetry. In contrast, $\mathit{M}$-point lasing allows deterministic control over emission channels via asymmetric pumping. We further show that spatially separated $\mathit{K}$-point lasers synchronize in phase and amplitude, undergoing SSB in lockstep. A theoretical density matrix approach cast into  stochastic differential equations reproduces the observed real- and Fourier-space intensity distributions and SSB behavior. Our findings establish spatially programmable metasurface lasers as a versatile platform for exploring 
dynamic phenomena in photonic lattices, with potential applications in vortex beam shaping, optical logic, and true random number generation.
\end{abstract}
\maketitle

\section{Introduction}
In recent years, metasurfaces have revolutionized the field of photonics, changing the paradigms of controlling light propagation and emission~\cite{kildishev2013planar,yu2014flat,vaskin_light-emitting_2019,schulz2024roadmap}. Metasurface lasers have attracted particular interest~\cite{kodigala_lasing_2017,ha_directional_2018,huang2020ultrafast,azzam2021single,mohamed2022controlling,arjas2024high,deng2025chiral,bashiri2025holographic}, expanding from previous developments in the field of distributed feedback (DFB) and photonic crystal lasers \cite{kogelnik_stimulated_1971, turnbull_relationship_2001,heliotis_emission_2004,notomi_directional_2001, imada_multidirectionally_2002, miyai_lasers_2006, liang_three-dimensional_2012}. 
In metasurface lasers, two-dimensional (2D) arrangements of nanostructures can be used to tailor not only the emission characteristics~\cite{huang2020ultrafast,azzam2021single,mohamed2022controlling,arjas2024high,deng2025chiral,bashiri2025holographic} but also the lasing process itself~\cite{wu_topological_2020,salerno_lossdriven_2022}. For example, plasmon lattice lasers take advantage of the strong feedback provided by high-quality factor diffractive modes and quasi-Bound State in the Continuum (quasi-BIC) conditions to achieve low-threshold lasing that is very robust to disorder \cite{stehr_low_2003, suh_plasmonic_2012, wang2017structural,guo_lasing_2019, heilmann_quasi-bic_2022, winkler_dual-wavelength_2020, schokker_lasing_2014, schokker_statistics_2015, schokker_systematic_2017, guo_spatial_2019, de_gaay_fortman_spontaneous_2024}. Although such lasers typically operate at the $\Gamma$-point, lasing at other high-symmetry points has also been extensively studied \cite{guo_lasing_2019, juarez_m-point_2022, wu_topological_2020, de_gaay_fortman_spontaneous_2024, de_gaay_fortman_accessing_2025}. Among plasmon lattice lasers, one of the most unique systems is a hexagonal lattice laser, in which the $\mathit{K}$ and $K'$ Bloch modes are time-reversed companions \cite{wu_topological_2020}, degenerate not only in frequency but also in the spatial mode profile. We have previously exploited this system as an experimental platform for studying spontaneous symmetry breaking (SSB)~\cite{de_gaay_fortman_spontaneous_2024}. SSB is a fundamental phenomenon in statistical physics,  
crucial, e.g., in understanding of phase transitions. On the other hand, photonic lattices with gain have received significant interest as a platform to emulate non-Hermitian Hamiltonians with controllable lattice symmetry, coupling strengths, and topology~\cite{kolkowski_pseudochirality_2021}. 

\begin{figure*}
    \centering
    \includegraphics[width=1\textwidth]{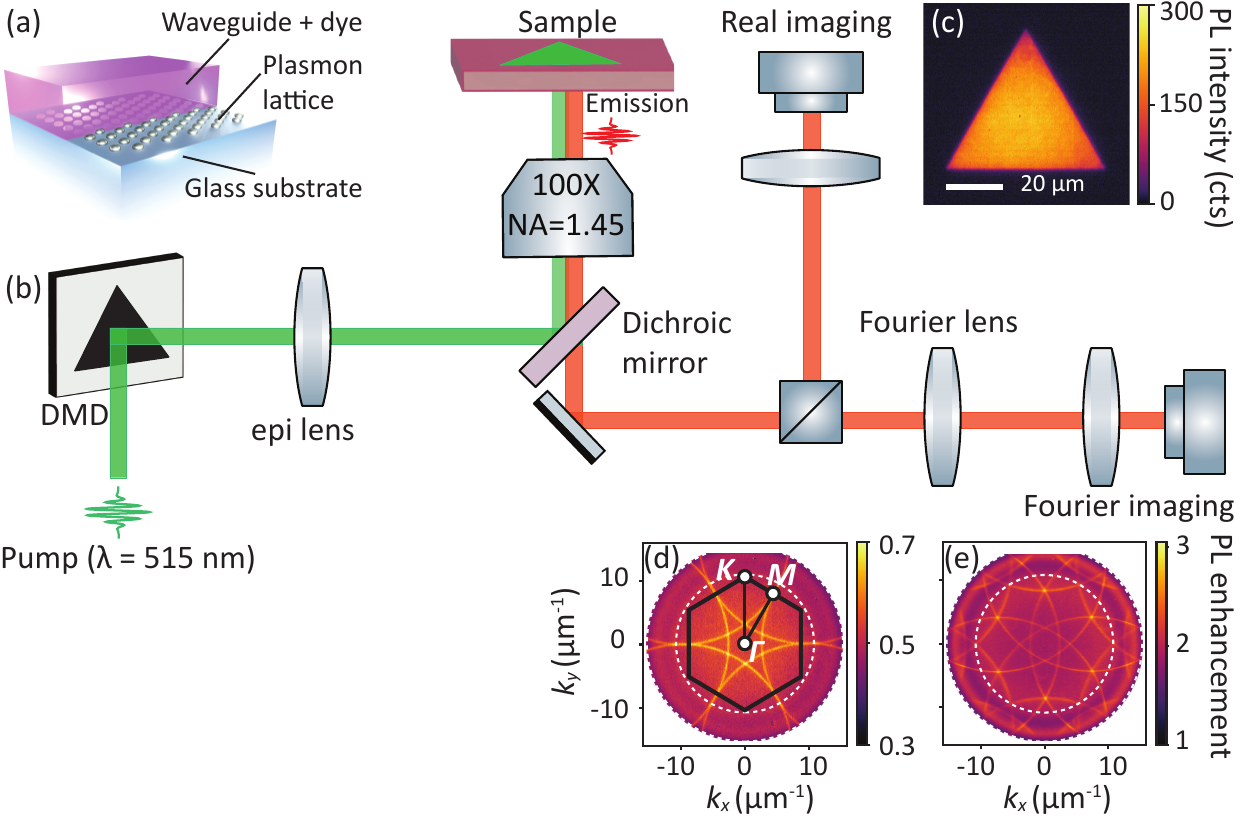}
    \caption{\textbf{Single-shot microscopy with structured pump illumination.} (a) Illustration of the plasmon lattice laser, which is a hexagonal lattice of silver nanoparticles embedded in a dye-doped slab waveguide. (b) Fluorescence microscope setup used to study plasmon lattice lasers. We place a digital micromirror device (DMD) in the pump-illumination track in the focus of the epi-lens, which enables us to project any pattern in the real-space plane of the sample. The emission signal is sent to the imaging setup where we record real- and Fourier-space images simultaneously. (c) 54 $\upmu$m sized triangle projected by the DMD revealed by fluorescence image recorded in real space. (d, e) Fourier images of fluorescence from the hexagonal plasmon lattices with pitches a = 375 nm and 500 nm, with bandpass filter centered at 580 nm wavelength, at which the repeated waveguide mode crossings occur at the $\mathit{M}$- and $\mathit{K}$-points, respectively. The first Brillouin zone is shown in black in panel (c), with high-symmetry points $\mathit{M}$, $\Gamma$, and $\mathit{K}$ indicated.}
    \label{fig:DMD_concept}
\end{figure*}

In metasurface lasers, the resonant cavity is spatially extended over the lattice, which creates an opportunity for controlling lasing by spatially structured pumping of the gain medium. To date, this additional degree of freedom has barely been explored. Here, we implement this idea in hexagonal plasmon lattice lasers, programming their spatial gain patterns using a digital micromirror device (DMD) that is placed in the pump beam. Unlike previous studies on finite-size DFB and photonic crystal lasers, which focused on systems with hard boundaries~\cite{kogelnik_coupledwave_1972,kazarinov_second-order_1985,sakai_two-dimensional_2007, liang_three-dimensional_2012, guo_spatial_2019, wang_lasing_2020, ren2022low}, our approach creates boundaries solely in the spatial distribution of gain across otherwise uniform periodic structures. We demonstrate the versatility of this approach by using it to realize $\mathit{K}$- and $\mathit{M}$-point lasing in plasmonic lattices. These systems operate on quasi-BICs as lasing modes, resulting in polarization vortex beams in the far-field \cite{imada_multidirectionally_2002, miyai_lasers_2006, liang_three-dimensional_2012, kodigala_lasing_2017, heilmann_quasi-bic_2022}. We investigate how controlling the spatial pump pattern determines the size and shape of the vortex beams. Moreover, as reported previously~\cite{de_gaay_fortman_spontaneous_2024}, $\mathit{K}/\mathit{K'}$-point lasing in hexagonal lattices simultaneously exhibits two types of SSB, namely, a parity-symmetry breaking (relative intensity between $\mathit{K}$ and $\mathit{K'}$ points), and a U(1) symmetry breaking (relative phase). An intriguing question is whether SSB persists when explicitly breaking the symmetry by a pump pattern that is incompatible with the underlying lattice and lasing mode symmetry. To this end, we address SSB in two distinct systems: SSB in $\mathit{K}/\mathit{K'}$-point lasers, entailing modes that are degenerate in energy and space, and SSB in $\mathit{M}$-point lasers, which entail modes degenerate in energy but not in spatial profile. These 
two classes of degeneracies at high-symmetry points present different levels of robustness against symmetry breaking induced by the pump beam. We juxtapose the exceptional robustness of $\mathit{K}$ points with the ability to dynamically bias SSB at the $\mathit{M}$ points by controlling the pump pattern. Finally, we demonstrate the potential of our methods for the study of separated nanolasers. Exciting questions arise regarding their coupling and synchronization: For instance, will nearby laser areas perform SSB independently, or in lockstep? And, if they operate in lockstep, will they synchronize in both amplitude and phase? Dynamically programming the gain area allows us to make coupled lattice lasers of chosen size, shape, and separation, providing a versatile platform to address these questions. We supplement our experimental results on lasing, SSB, and synchronization, with a density-matrix based model which takes into account the gain medium dynamics, programmed spatial distribution of the pumping rate,  random spontaneous emission as seed, and the symmetry of the $\mathit{K}$/$\mathit{K'}$ Bloch modes. The obtained theoretical results provide additional insights into the unique physics of lattice lasers under structured pumping.

\section{Experimental approach}
To create gain regions with sharply defined and controllable boundaries, we structure the pump beam using a digital micromirror device (DMD). This method is borrowed from programmable emission control in polariton condensate studies \cite{cristofolini_optical_2013, schmutzler_all-optical_2015}, planar random lasers \cite{bachelard_adaptive_2014}, and luminescent all-dielectric metasurfaces \cite{iyer_sub-picosecond_2023}. Figure~\ref{fig:DMD_concept}(a) shows the structure of our samples, while Fig.~\ref{fig:DMD_concept}(b) illustrates the operation principle of our setup, which is essentially a fluorescence microscope driven by a 250 fs pulsed pump laser of 515 nm wavelength (LightConversion Pharos, frequency doubled).
First, the pump beam is reflected by a Texas Instruments digital mirror device (DLP6500, a 2 megapixel device with 7.6 $\upmu$m pixel size). Next, we image the structured pump beam through the microscope epi-illumination lens and objective ($\text{NA} = 1.45$ oil objective, Nikon CFI Plan Apochromat lambda 100$\times$) onto the sample, at a demagnification of 100 times. This allows us to project an amplitude mask formed by DMD pixel-mirrors as intensity pattern in the pump field on the sample with diffraction-limited resolution $\lambda/2\text{NA}=175$ nm ($\approx2.5$ DMD pixels). Figure~\ref{fig:DMD_concept}(c) shows an example of real-space illumination, visualized by recording fluorescence from a dye-doped polymer slab (without plasmonic nanoparticles). The emission clearly delineates a triangular shape, digitally encoded by the DMD onto the pump. As in Ref. \cite{de_gaay_fortman_spontaneous_2024}, our setup can operate in single-shot mode, with a camera frame rate synchronized to the laser repetition rate (set at 20 Hz). A beamsplitter in the imaging track ensures simultaneous capture of real-space images with one camera and Fourier-space images with the other. With single-shot recordings, we capture shot-to-shot variation in the laser emission, originating, e.g., from mode competition and SSB \cite{keitel_single-pulse_2021, de_gaay_fortman_spontaneous_2024}. 

We study lattice lasers composed of hexagonal arrays of plasmonic nanoparticles, sketched in Fig.~\ref{fig:DMD_concept}(a). The nanoparticles are 80 nm in diameter and 35 nm in height, and the lattices have a pitch of 375 nm or 500 nm (designed for either $\mathit{M}$ or $\mathit{K}$-point lasing, respectively). The arrays are fabricated by standard electron beam lithography followed by lift-off (nanofabrication protocol identical to that reported in Refs.~\cite{de_gaay_fortman_spontaneous_2024, de_gaay_fortman_accessing_2025}). We cover the arrays with an approximately 500 nm thick SU8 polymer slab doped with 3.5 wt \% rhodamine 6G providing the gain required for lasing. The nanoparticle geometry and waveguide thickness are chosen such that the in-plane-polarized localized surface plasmon resonance of the nanoparticles overlaps spectrally with the gain peak (near 590 nm), and the waveguide supports only one transverse electric (TE) and one transverse magnetic (TM) guided mode. The in-plane-polarized electric field of the TE mode is aligned with the main polarizability axes of the disk-shaped nanoparticles, favoring diffractive coupling through the TE mode over the TM mode \cite{schokker_lasing_2014, schokker_systematic_2017}. 

Figures \ref{fig:DMD_concept}(d, e) show Fourier images of the photoluminescence (PL) enhancement from plasmon lattices with pitches 375 and 500 nm. In the extended-zone scheme representation of band-structure physics, the waveguide mode represents a circular feature in the $(k_x,k_y)$ space which reflects the continuous rotational symmetry of the slab. The circles are repeated at each reciprocal lattice point of the hexagonal lattice \cite{vaskin_light-emitting_2019}. This repeated-zone-scheme dispersion is immediately evident in the back-focal-plane images in Fig.~\ref{fig:DMD_concept}(d) and (e). The first Brillouin zone (BZ) is hexagonally shaped and is highlighted in black in Fig.~\ref{fig:DMD_concept}(d), with high-symmetry points $\Gamma$ (center of BZ), $\mathit{K}$ (corners of BZ), and $\mathit{M}$ (in between corners of BZ). In Fig.~\ref{fig:DMD_concept}(d), for $a=375$ nm, the repeated waveguide modes intersect at the $\mathit{M}$-points, while in Fig.\ref{fig:DMD_concept}(e) for $a=500$ nm, the intersections occur at the $\mathit{K}$-points. At such intersections, two or more Bloch modes couple and form a stop gap in $(\omega,k_{||})$ space, with band edges suitable for lasing \cite{kogelnik_coupledwave_1972, notomi_directional_2001, schokker_lasing_2014, guo_spatial_2019}. For the plasmon lattice of pitch 500 nm, lasing occurs on the $A_1$ mode at the $\mathit{K}$-point \cite{ochiai_dispersion_2001, wheeldon_symmetry_2007, malterre_symmetry_2011, guo_lasing_2019, de_gaay_fortman_spontaneous_2024}. In all of the experiments reported in this work, we vary boundary conditions only  by shaping the excitation area, not the hard boundaries of the lattice. The total size of each fabricated  lattice is 200~$\upmu$m $\times$ 200~$\upmu$m, much larger than the projected pump fields. Details on our measurement protocol can be found in the SI.

\section{$\mathit{K}$-point lasing of individual gain areas}
\begin{figure*}
    \centering
    \includegraphics[width=1\textwidth]{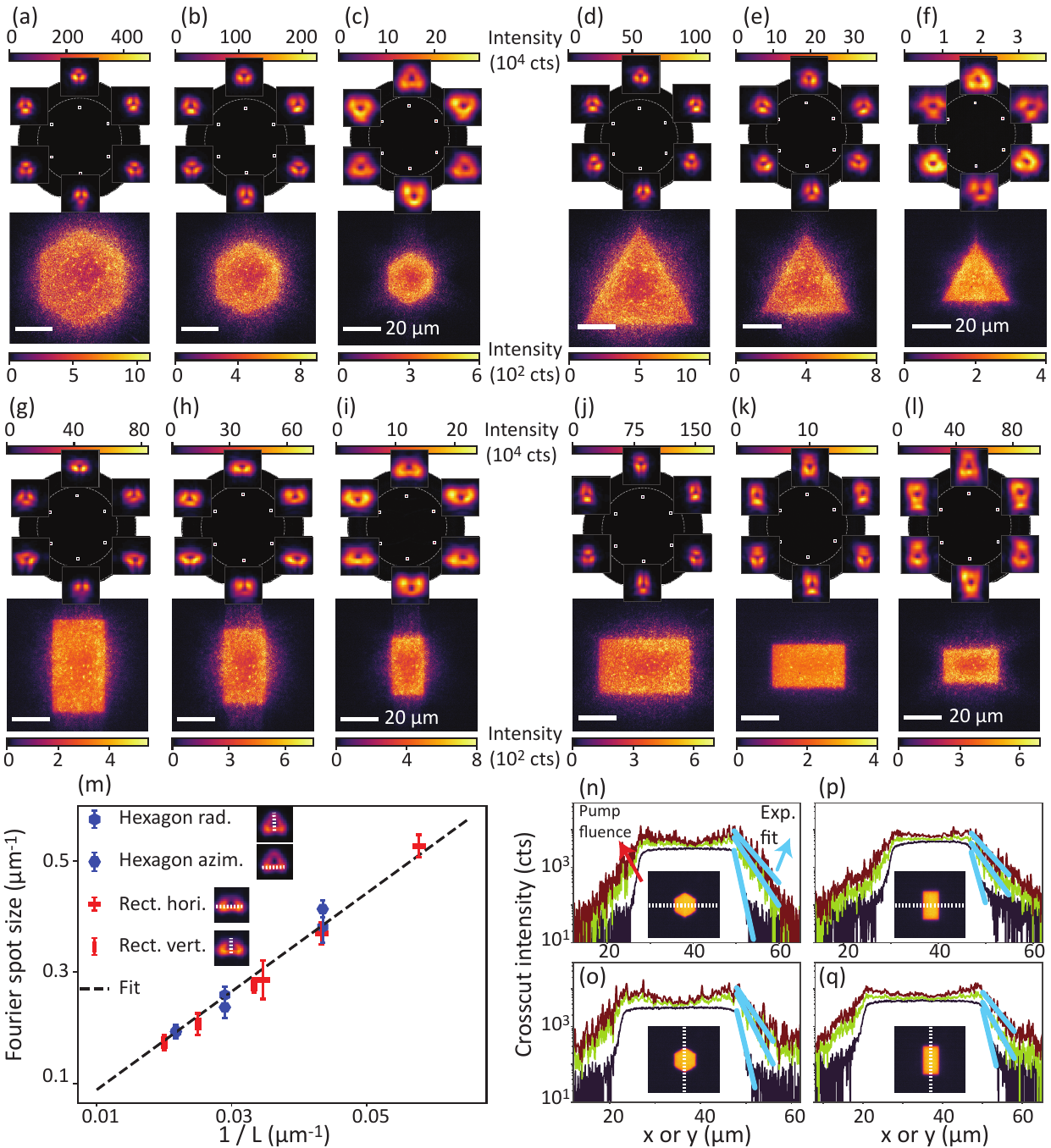}
    \caption{\textbf{$\mathit{K}$-point lasing with boundary conditions in gain, averaged over many lasing shots.} (a-c) Fourier- and real-space images of $\mathit{K}$-point lasing ($a=500$ nm) from DMD-projected hexagons of short diagonal $L=\{46,34.5,23\}$ $\upmu$m (radius $\{26.6,19.9,13.3\}$ $\upmu$m). (d-f) Lasing from triangles with side length $L=\{57.5,46,34.5\}$ $\upmu$m. (g-i) Lasing from rectangles of shape $\{(28.9\times50), (23.1\times40),(17.3\times30)\}$ $\upmu$m and (j-l) the same rectangles rotated by 90$\degree$. In real space, lasing is characterized by finely spaced speckles, overlayed with intensity envelopes that each have a minimum at the center. 
    In Fourier space, six $\mathit{K}$-point lasing spots are donut-shaped (see insets; the extent of each inset is $\Delta k/k_0=0.06$) but relate to the real-space laser shapes and sizes by Fourier transform, increasing in their Fourier-space extent as the real-space area is reduced. (m) Fourier-space lasing spot size versus the inverse of real-space length $L$. This data is obtained by measuring the Fourier-space extent of the spots, averaging over six spots and taking standard deviation for error estimation. In blue, we display the results for radial and azimuthal crosscuts through donut-spots 
    for the hexagon pump (panels a-c), whereas red points show the results for horizontal and vertical crosscuts through donut-spots for the rectangle pump (panels g-i). Both datasets are plotted against 1/$L$ with $L$ being either the hexagon's short diagonal or the relevant rectangle's side length. The linear relation $\Delta k \propto 1/L$ clearly holds.  
    \textit{Caption continues on the next page}.}
    \label{fig:multi_shot}  
\end{figure*}

\addtocounter{figure}{-1}
\begin{figure} [t!]
  \caption{\textit{Continuation of caption from the previous page.} (n,o) To quantify the propagation length of laser light beyond the pumped area, we take horizontal and linear crosscuts (at $y_c,x_c=(35,38.7)$ $\upmu$m and summing 20 pixels wide) through the real-space hexagon shape in panel (c), for pump fluences $\{1.5,2.1,2.7 \}$ mJ/cm$^2$. The lowest fluence (dark purple curve) belongs to fluorescence, the higher two (green/red curves) to lasing. We fit the line $y = a\exp{b(x-x_1)}$ (cyan curves) to the region beyond the pumped area and obtain $b = -1.2$ $\upmu$m$^{-1}$ for fluorescence and $b = -0.4$ $\upmu$m$^{-1}$ for lasing. The laser emission thus propagates about 2 to 3 $\upmu$m beyond the gain boundary. (p,q) Analogous crosscut results for the rectangular pump 
  (panel (i)), with fluences $\{1.8,2.3,2.8 \}$ mJ/cm$^2$ -- fitting leads to similar $b$-values as for the hexagons.}
\end{figure}

Figure~\ref{fig:multi_shot} displays real- and Fourier-space images of laser emission for pumped areas of different shapes and sizes. Each image is obtained by averaging over 100 consecutive single-shots, showing behavior of the $\mathit{K}$-point lasing output well above the lasing threshold. Specifically, in Fig.~\ref{fig:multi_shot}, we show real-space and corresponding Fourier-space lasing images for hexagonal (a-c), triangular (d-f) and rectangular (g-l) areas of decreasing size. Areas of smaller sizes are found to exhibit higher lasing thresholds due to less feedback, as well as increased radiative loss and side leakage. In each of the real-space images, we observe an intensity envelope that peaks at the edges of the lasing area, and reaches a minimum in its center. In the theory of 1D distributed feedback lasers by Kogelnik and Shank~\cite{kogelnik_coupledwave_1972}, such profiles are viewed as a clear signature of an undercoupled lasing system, which is typical for photonic and plasmonic lattice lasers \cite{sakai_two-dimensional_2007, liang_three-dimensional_2012, guo_spatial_2019}. Since lasing preferentially selects a mode with the least (radiative and absorptive) loss, lattice lasing is associated with dark-mode~\cite{hakala_lasing_2017} or quasi-BIC~\cite{kodigala_lasing_2017, zhang_observation_2018, ha_directional_2018, heilmann_quasi-bic_2022,mohamed2022controlling}, which entails Fourier-space output in the form of a donut-shaped intensity spot that displays a minimum in its center \cite{liang_three-dimensional_2012}. Indeed, this is what we generally observe in all of our Fourier images of lasing emission (see the SI for more examples), in line with many other 2D DFB lasing studies \cite{imada_multidirectionally_2002, heliotis_emission_2004, miyai_lasers_2006, liang_three-dimensional_2012, tenner_two-mode_2018}. 
In Fig.~\ref{fig:multi_shot}, the lasing occurs on a stop-gap edge at the $\mathit{K}$-point that corresponds to radiation-forbidden Bloch modes. However, the finite extent of the lasing area opens up the possibility to radiate, manifesting itself as a ring of emission around the $\mathit{K}$-point, i.e., the observed donut-shaped intensity spots, known to exhibit a polarization vortex \cite{miyai_lasers_2006, zhang_observation_2018, doeleman_experimental_2018, tenner_two-mode_2018, heilmann_quasi-bic_2022}.

Apart from the quasi-BIC character evident in far-field radiation, we observe that the vortex beams are structured by the real-space pump pattern. For example, Fig.~\ref{fig:multi_shot} shows that, in Fourier space, the laser spots become larger as the real-space gain area gets smaller. Figures~\ref{fig:multi_shot}(g-l) show that, for rectangular pump patterns, the resulting Fourier-space donuts also become rectangularly shaped, but with their long axis oriented perpendicular to the long axis of the rectangle in real space. This behavior follows the reasoning that far-field emission is the Fourier transform of the field at the sample plane. The real space can qualitatively be seen as the product of the infinite-system Bloch mode and the clip imparted by the finite lasing area. According to this logic, far-field output is the convolution of the infinite-system output with the Fourier transform of the lasing area \cite{liang_three-dimensional_2012, tenner_two-mode_2018, guo_spatial_2019}. Indeed, we find that the Fourier-space width $\Delta k$ of the lasing spots is inversely  proportional to the laser size $L$, as analyzed in Fig.~\ref{fig:multi_shot}(m).

Finally, the real-space images show that some lasing emission is present just outside the pumped areas. While the DMD provides a sharp boundary between excited and non-excited areas (see Fig~\ref{fig:DMD_concept}(c)), there is no interruption for Bloch wave propagation as the metasurface continues beyond this boundary. In the non-excited regions, the Bloch mode travels in the metasurface along the $\mathit{K}$-direction without gain, loosing its energy by Bragg reflection, diffractive outcoupling, and absorption. To reveal this behavior, we plot in Figs.~\ref{fig:multi_shot}(n,o) and (p,q) the real-space intensity profiles across the illuminated areas for increasing pump fluences. Clearly, intensity persists beyond the gain boundary, where it decays exponentially. The lasing intensity (green and red curves) decays more slowly than the fluorescence intensity (dark purple curve), which we take as a measure for how sharply the gain region is defined. From exponential fits (cyan curves, linear on the log scale) we obtain decay constants of $1.2$ $\upmu$m$^{-1}$ for  fluorescence intensity, and $0.4$ $\upmu$m$^{-1}$ for laser intensity. This indicates that the lasing emission propagates about 2-3 $\upmu$m away from the pumped area, corresponding to 6-10 unit cells. In the theory of photonic crystals, an important length scale is the so-called `Bragg length', which is a metric for the number of unit cells required to build up a significant Bragg reflection. For DFB lasers, this length indicates the distance required to build up a significant optical feedback. The measured 6-10 unit cells length is consistent with reported minimum sizes required for lasing in plasmon lattice arrays of finite extent (hard boundaries)~\cite{guo_spatial_2019, wang_lasing_2020}.

\section{Spontaneous symmetry breaking \\in $\mathit{K}$-point lasers}
\begin{figure*}
    \centering
    \includegraphics[width=1\textwidth]{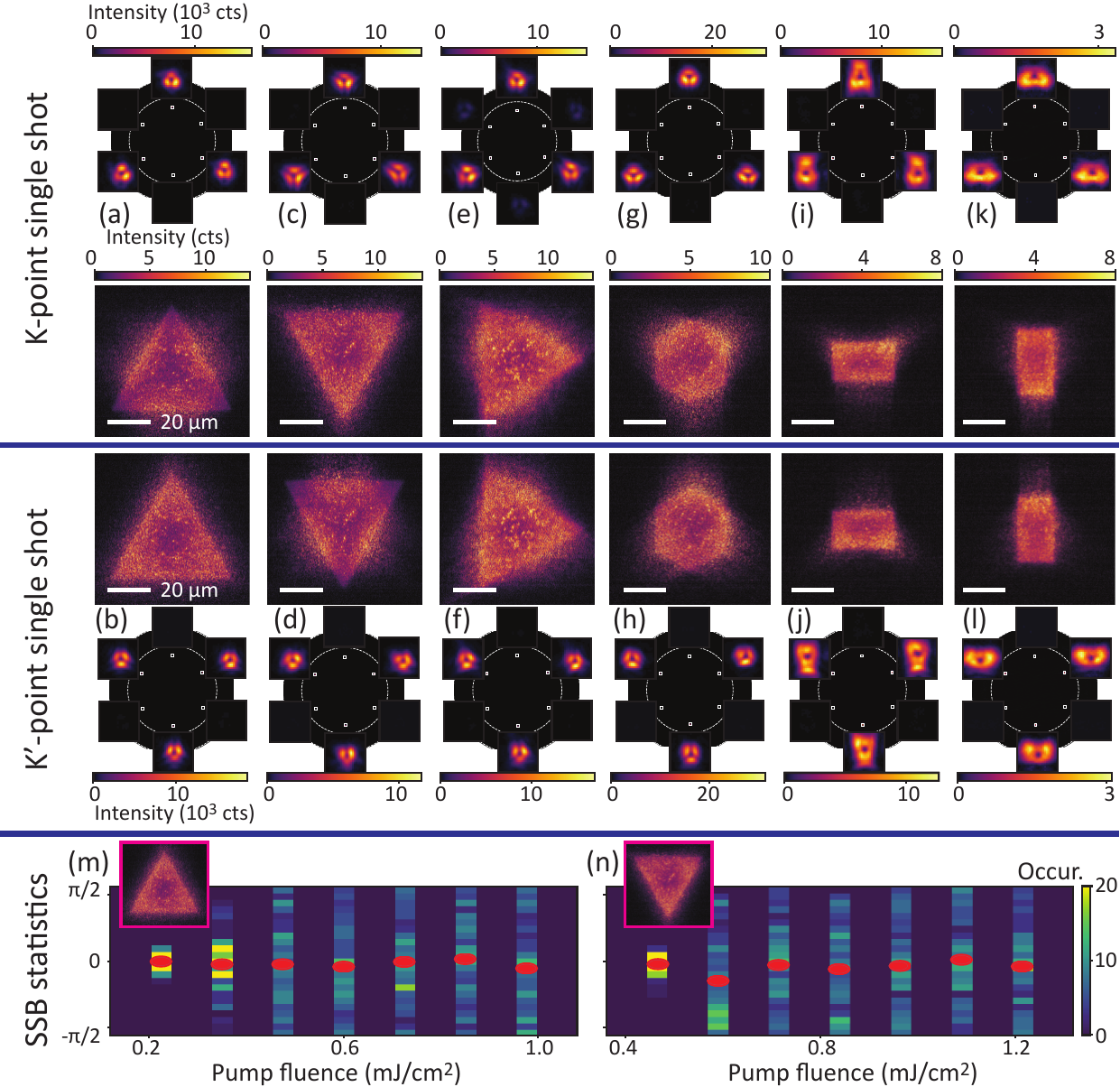}
    \caption{\textbf{Single-shot $\mathit{K}$-point lasing and SSB.} Panels (a,c,e,g,i,k): Simultaneously recorded Fourier-space (top) real-space (bottom) single shots, in which SSB leads to dominant $\mathit{K}$-mode lasing, for a variety of lasing area shapes. For each shape, in addition to displaying the intensity dip in the center, the real-space envelope forms three high-intensity lobes at the edges of the lasing area, forming an equilateral triangle. The orientation of that triangular envelope pattern is mirrored with respect to the $\mathit{K}$ direction of the lasing mode in Fourier space. At the same time, the real-space envelope is directly imprinted on the donut-shaped emission spots. Panels (b,d,f,h,j,l): Single shots for the same pump patterns but with SSB giving rise to dominant $\mathit{K'}$-mode lasing. Pump patterns: (a-f) Triangle, side length 53 $\upmu$m. (g,h) Hexagon, short diagonal 34.5 $\upmu$m. (i-l) Rectangle, shape $(17.3\times30)$ $\upmu$m. (m) Single-shot statistics of $\mathit{K}/\mathit{K'}$ SSB parameter $\theta$ that traces relative amplitude, as a function of pump fluence for the upward triangle (panels a,b), and (n) for the downward triangle (panels c,d). Note the absence of bias towards $\mathit{K}/\mathit{K'}$ lasing, despite breaking their symmetry by the gain distribution.}
    \label{fig:single_shot}
\end{figure*}

In hexagonal lattices, the $\mathit{K}$- and $\mathit{K'}$-point modes are uncoupled, but display a unique type of degeneracies. Within each of the modes ($\mathit{K}$ and $\mathit{K'}$), three wave vectors are coupled by lattice diffraction. These $\mathit{K}$ and $\mathit{K'}$ 
triplets are each other's time reverse, which makes them degenerate both in eigenfrequency and in near-field intensity distribution. These unique symmetry properties of the $\mathit{K}$ and $\mathit{K'}$ modes allowed them to be used to observe spontaneous symmetry breaking (SSB)~\cite{de_gaay_fortman_spontaneous_2024}. In SSB studies, an important question is how SSB occurs as a function of boundary conditions, which may be explicitly chosen to break the symmetry of an otherwise infinite system. To understand how $\mathit{K}$-point lasing and the associated SSB behave as a function of gain boundary conditions, we perform single-shot measurements for pumped areas of various shapes. For increasing pump fluence, we record sequences of $\geq 100$ single shots. As in Ref. \cite{de_gaay_fortman_spontaneous_2024}, we observe spontaneous breaking of parity symmetry between the intensities of the two degenerate modes, one lasing at the $\mathit{K}$ points, and another at the $\mathit{K'}$ points.  While \textit{within} a triplet of $\mathit{K}$ (resp. $\mathit{K'}$) points, the relative intensities are equal, the relative intensity \textit{between} the $\mathit{K}$- and $\mathit{K'}$-point triplets is random in every shot.

Selected shots of high-contrast $\mathit{K}$-mode lasing for various shapes are shown in the top row of Figure~\ref{fig:single_shot}, while the corresponding $\mathit{K}'$-mode lasing shots are shown in the bottom row. Each example is well above the lasing threshold.  Independently of the shape of the pumped area, we observe that the real-space intensity envelope is directly linked to the triangular Fourier-space output of the laser: The real-space intensity envelopes create triangular patterns, in which the bright spots (triangle vertices) are in locations exactly opposite to the lasing mode wave vectors. Figures~\ref{fig:single_shot}(a,b,g,h) show this for triangularly shaped lasers. For $\mathit{K}$-point lasing in Fig.~\ref{fig:single_shot}(a), the real-space envelopes are brightest at the sides of the triangular lasing area,
whereas, \textit{vice versa}, for $\mathit{K'}$-point lasing in Fig.~\ref{fig:single_shot}(b), the envelope has the highest intensity at the corners of the lasing area. This behavior is not affected by the geometry of the pumped area. For example, the hexagonal gain area in (d,j) shows the same behavior, even though this geometry is fully symmetric with respect to $\mathit{K}/\mathit{K'}$ lasing in Fourier space. This behavior shows that the wave vectors of the involved Bloch modes, mapped in Fourier-space as a triangle of output spots, are oppositely oriented to the in-plane directions, in which the intensity builds up upon propagation through the amplifying regions of the metasurface.  

Our experiment addresses the important question of whether boundary conditions in gain might bias the SSB in the relative intensity of the lasing modes. Fig.~\ref{fig:single_shot}(m,n) shows the recorded statistical spread of the $\mathit{K}$ and $\mathit{K'}$ intensities. We express the $\mathit{K}/\mathit{K'}$ intensity ratio with the parameter $\theta$, defined as $\theta = 2\arctan(\frac{I_K - I_{\mathit{K'}}}{I_K + I_{\mathit{K'}}})$, where $I_K$ and $I_{\mathit{K'}}$ are the intensities of the output spots summed separately for the $\mathit{K}$ and $\mathit{K'}$ points. This parameter is computed for every shot in the time trace. We do not observe any bias towards $\mathit{K}$- or $\mathit{K'}$-point lasing, despite the fact that we explicitly break the $\mathit{K}/\mathit{K'}$ parity symmetry as we align triangular real-space gain areas with the orientation of either $\mathit{K}$ or $\mathit{K'}$ wave vector. The same lack of bias is observed for all other recorded examples of pumping geometries, as shown in the SI. In Ref. \cite{de_gaay_fortman_spontaneous_2024}, we argued that, as the $\mathit{K}$ and $\mathit{K'}$ modes are each other's time-reverse counterparts and thus share identical real-space amplitude profiles, they are inherently robust against symmetry breaking imposed by asymmetric pumping, regardless of the shape of the pumped area. The experimentally observed  real-space envelopes, as well as the unique robustness of the $\mathit{K}/\mathit{K'}$ SSB behavior, are reproduced by our theoretical model presented in the final section of this paper.

\begin{figure*}
    \centering
    \includegraphics[width=1\textwidth]{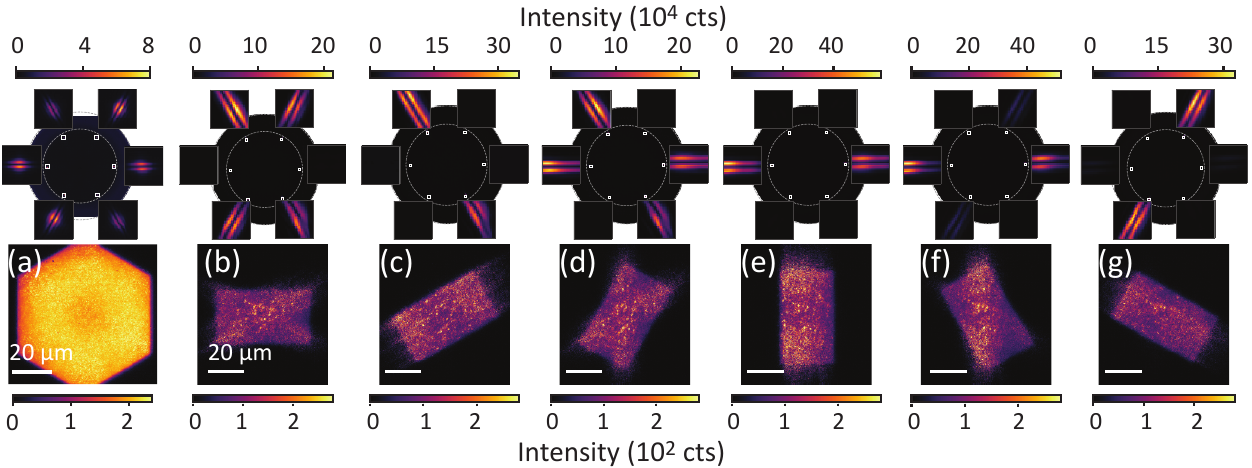}
    \caption{\textbf{DMD-controlled $\mathit{M}$-point lasing.} (a) 20-shot average Fourier- and real-space images of lasing at the $\mathit{M}$ points under symmetric pumping (LCP-polarized pump forming hexagonal shape in real space with short diagonal 69 $\upmu$m). SSB between the three frequency-degenerate $\mathit{M}$ modes leads to intensity fluctuations of $15\%$. (b-g) Active switching between $\mathit{M}$ modes by rotating a rectangular pump shape $(50\times28.9)$ $\upmu$m. The bias is achieved by overlapping the gain pattern with the spatially distinct mode profiles belonging to one (c,e,g) or two (b,d,f) of the three $\mathit{M}$ modes. The insets magnifying the lasing spots in the Fourier images have extents $k_{||}/k_0=0.1$.}
    \label{fig:M_point}
\end{figure*}

\newpage

\,
\newpage

\,
\newpage 

\section{On-demand symmetry breaking \\in $\mathit{M}$-point lasers}

The property that SSB cannot be biased by boundary conditions in programmable gain geometries, established from Figure~\ref{fig:single_shot}, is specific for $\mathit{K}/\mathit{K'}$-point lasing. As said, this robustness is intrinsic to their time-reversal relation, which uniquely implies degeneracy in both energy ($\hbar \omega$) and space ($|E(r)|^2$), due to which spatial amplitude perturbations project equally on both modes. To expose the unique nature of $\mathit{K}/\mathit{K'}$-point lasing, we compare it with experiments on $\mathit{M}$-point lasers. When the lattice pitch is set to 375 nm, the $\mathit{M}$-point band intersections fall within the rhodamine 6G gain window (see Fig.~\ref{fig:DMD_concept}(d)). Similarly to $\mathit{K}$ points, the first BZ contains six $\mathit{M}$ points, although their symmetry properties are markedly distinct from those of the $\mathit{K}$ points: Each of the three modes engages two diametrically opposite $\mathit{M}$ points that connect through a reciprocal lattice vector. This is in contrast to the $\mathit{K/K'}$ modes, where each of the two modes engages three $\mathit{K/K'}$ points. Under fully symmetric gain conditions, mode competition between the three $\mathit{M}$ modes occurs, leading to SSB in relative intensity. We show this in Fig.~\ref{fig:M_point}(a), where a left-handed circularly polarized (LCP) pump, shaped into a hexagon, creates a Fourier space laser output that is, on average, almost completely symmetrically distributed over the three modes. SSB in intensity is evident from shot-to-shot relative intensity variations of about $15 \%$ between the $\mathit{M}$-point pairs. 

The stark contrast to $\mathit{K}/\mathit{K'}$-point lasing is evident from the fact that one can easily bias the system towards lasing from one of the $\mathit{M}$ modes using an elongated (e.g., rectangular) pump pattern oriented perpendicular to the pair of selected $\mathit{M}$ points. The bias results from the fact that the real-space mode profiles of the $\mathit{M}$ points are not degenerate, as opposed to the $\mathit{K}/\mathit{K'}$ points. Instead, the $\mathit{M}$ modes each form a simple one-dimensional standing wave pattern normal to their specific $\mathit{M}$ direction. This one-dimensional nature is evident from the stripe-shaped lasing output in the Fourier images, a typical output of 1D DFB lasers.  Figure~\ref{fig:M_point}(b-g) shows the results for a setting in which the pump is still LCP-polarized, but shaped to produce a rectangular gain area, the long axis of which is rotated in steps of 30 degrees. This ensures mode overlap with two $\mathit{M}$-point modes (panels b,d,f), or with a single $\mathit{M}$ mode (c,e,g). As a consequence, the system is indeed forced to either show SSB between two modes, or lasing on just a single $\mathit{M}$ mode. Notably, we found that another way to bias $\mathit{M}$-mode lasing is to project a fully symmetric pump pattern while changing the angle of its linear polarization. This does not change the laser boundary, but instead modifies the pump field distribution within each unit cell, affecting especially the orientation of electric-field hot spots at the surface of plasmonic nanoparticles. The above findings demonstrate the unique capabilities of our structured pumping approach to explore the symmetry properties of diverse photonic lattices and the potential to dynamically control their lasing emission.

\section{Synchronization of lasing and SSB in coupled metasurface lasers}
\begin{figure*}
    \centering
    \includegraphics[width=1\textwidth]{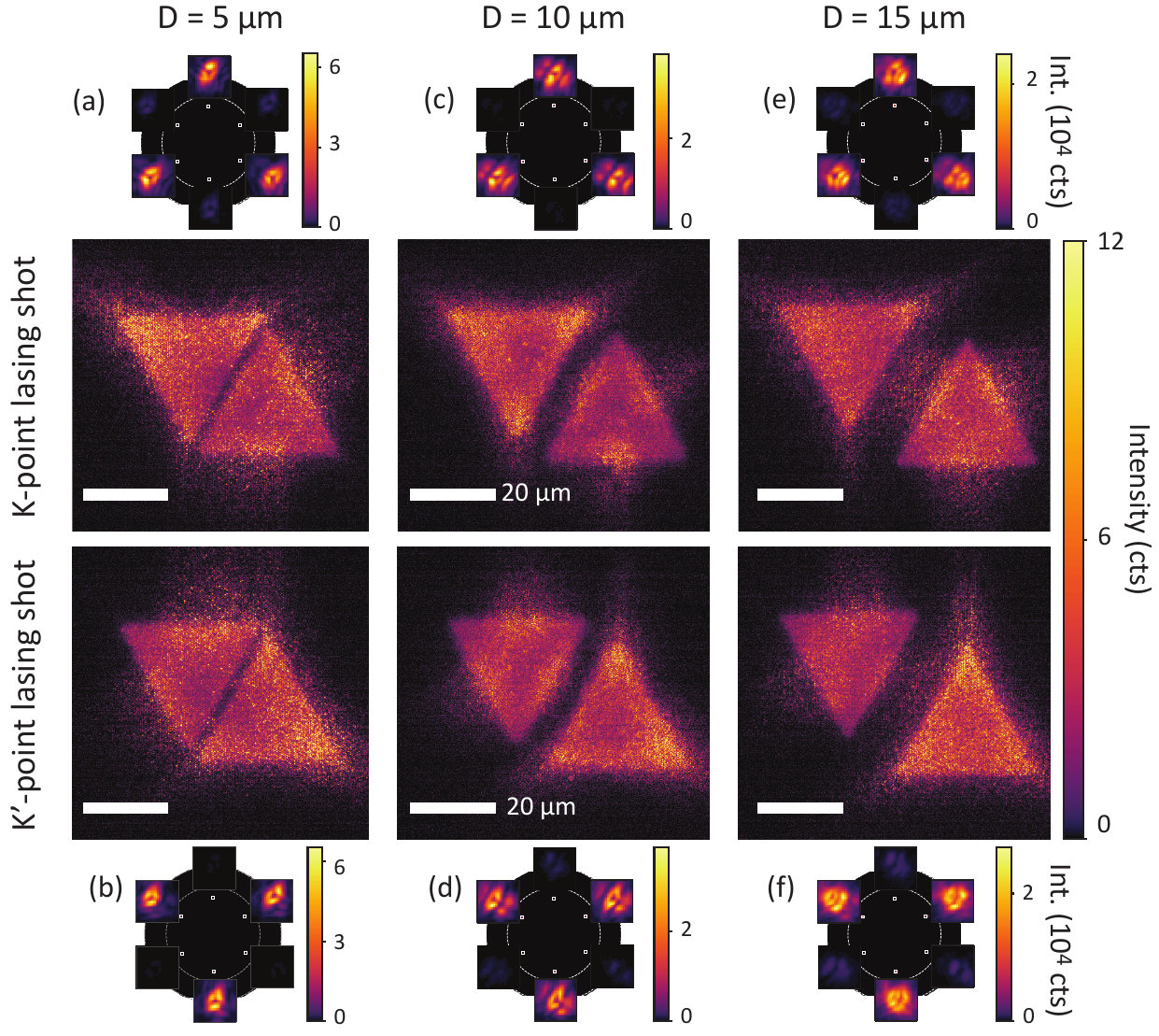}
    \caption{\textbf{$\mathit{K}$- and $\mathit{K'}$-point lasing in pairs of separated lasing areas.}
    (a-c) Examples of simultaneously recorded single-shot Fourier and real-space images of $\mathit{K}$-point lasing in two separated triangular lasing areas as a function of their separation distance (D). (d-f) Analogous examples of $\mathit{K'}$-point lasing. Each of the triangular pump patterns has 34.5 $\upmu$m side length. In the Fourier images, fringes in the lasing spots indicate phase synchronization between the two lasing areas, whereas the real-space images reveal nontrivial intensity distributions that change upon increasing D. The intensity patterns associated with $\mathit{K}$- and $\mathit{K'}$-mode lasing appear to be each other's mirror images. The Fourier image insets extend over $k_{||}/k_0=0.06$.}
    \label{fig:coupled_triangles}
\end{figure*}

The study of coupling between separated metasurfaces as a function of their shape and distance is another exciting possibility offered by spatially programmable pump patterns. To explore this possibility, we investigate $\mathit{K}$-point lasing in pairs of closely spaced triangular gain areas, each with side length 33 $\upmu$m. The two lasing areas can couple via Bloch modes propagating through the unpumped region that separates them. As before, we record many single-shot lasing sequences to observe SSB. Representative events of high-contrast $\mathit{K}$-point lasing are presented in the top row of Fig.~\ref{fig:coupled_triangles}, and those of $\mathit{K'}$-point lasing in the bottom row. From left to right, we increase the separation distance D by moving the left triangle up-left and the right triangle down-right, both along the $\mathit{K}$ direction. Figure ~\ref{fig:coupled_triangles}(a) (D = 5~$\upmu$m) shows that the $\mathit{K}$-point lasing spots are elongated along the direction perpendicular to the diagonal spanned by the two triangles. In the real-space, the mode envelope forms a supermode with the highest intensity at the edges of the right triangle and at the corners of the left triangle. In addition, the supermode shows an intensity dip in the center of the total lasing area. Such an intensity envelope and Fourier output imply that the two lasing areas essentially form a single, individual undercoupled laser, with the shape of the enclosing parallelogram. For high-contrast $\mathit{K'}$-mode lasing in Fig.~\ref{fig:coupled_triangles}(b), the real-space intensity pattern is mirrored, in accordance with the parity symmetry between the $\mathit{K}$ and $\mathit{K'}$ modes. Next, as we increase the separation distance to D = 10~$\upmu$m (see Fig.~\ref{fig:coupled_triangles}(c,d)), we can observe interference fringes that appear in the $\mathit{K}$-point donut spots along the separation direction. The fringes indicate that the two lasers are mutually coherent, allowing their emission to interfere in the far field. Within a 100-long single-shot time trace, the fringe pattern in each of the 6 output beams hardly varies from shot to shot, which indicates that the relative phase at which both lasers radiate is locked: Although the system exhibits $\mathit{K}$/$\mathit{K'}$-mode SSB, the two lasers undergo SSB in lockstep. We further substantiate this claim in the SI by measuring the relative phase between the $\mathit{K}/\mathit{K'}$ lasing modes at the edge of each real-space triangular pattern. Extracting the relative phase is made possible by resolving the real-space intensity distribution in the lasing area within the limit of our imaging system and by correlating it with the nanoparticle lattice (as described in the supplement of Ref.~\cite{de_gaay_fortman_spontaneous_2024}). In addition to the interference fringes in the Fourier space, the real-space images for D = 10~$\upmu$m in Fig.~\ref{fig:coupled_triangles}(c,d) turn out to exhibit significantly different $\mathit{K}/\mathit{K'}$-point 
supermodes compared to the D = 5~$\upmu$m case (panels a,b). For D = 10~$\upmu$m, each shape clearly forms its own spatial intensity distribution with a dip in the center of each triangle, suggesting that the system behaves as two separate undercoupled DFB lasers, synchronized through coupling mediated by the $\mathit{K/K'}$ Bloch modes propagating across the gap. Finally, for the largest separation (D = 15~$\upmu$m, see Fig.~\ref{fig:coupled_triangles} (c) and (f)), the donut-shaped output spots in Fourier space have more closely spaced fringes, consistent with the larger laser separation. The fringe contrast is lower than for D = 10~$\upmu$m, which might be due to the limited Fourier-space resolution. On the other hand, this lower contrast could also be attributed to a lower mutual coherence between the two lasers due to an increased coupling distance. 

\begin{figure*}
    \centering
    \includegraphics[width=1\textwidth]{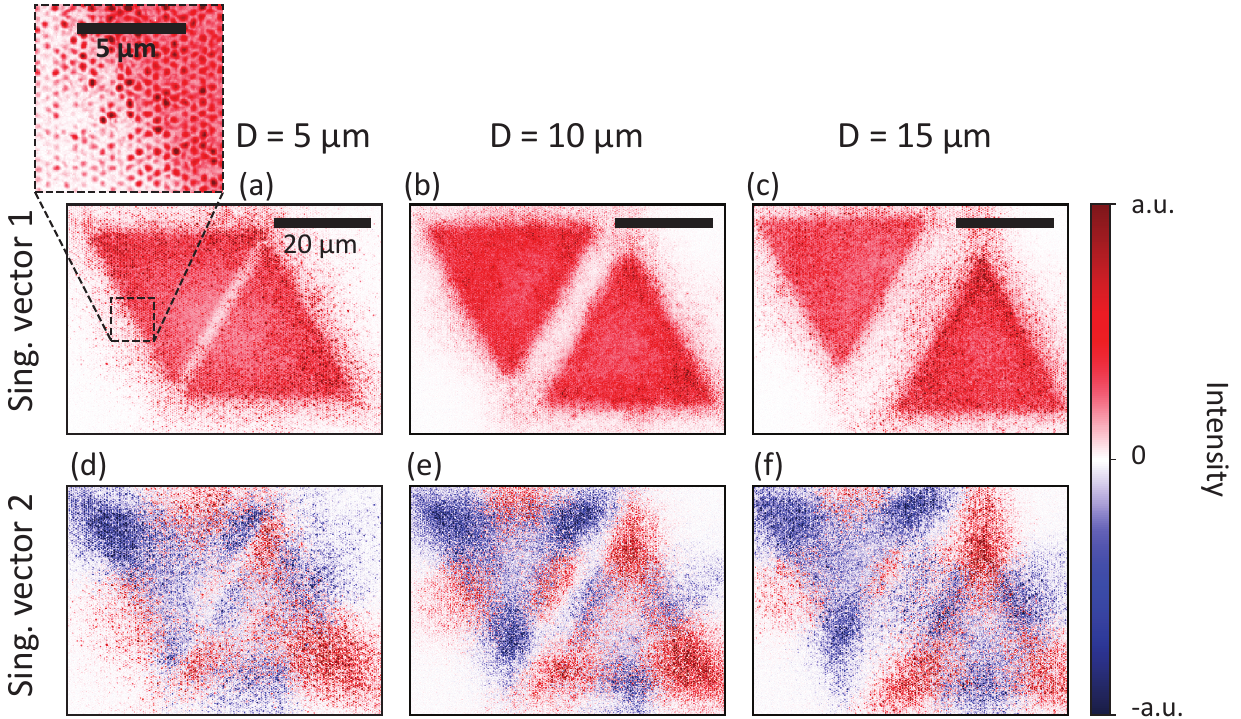}
    \caption{\textbf{Revealing the SSB synchronization by singular value decomposition (SVD) for two coupled lasers.} (a-c) First singular vectors, obtained by SVD analysis of 100 single-shot real-space images for increasing separation distance D between the two triangular lasers. These vectors correspond to the intensity distribution averaged over the sequence of images. Note in the inset how the intensity distribution reveals the underlying hexagonal lattice. In (a), the real-space envelope extends over the full parallelogram formed by the two closely spaced lasers. In (b) and (c), on the other hand, the separation is large enough for the two shapes to come with their own undercoupled envelopes (intensity dip in center of each triangle). (d-f) Second singular vectors, showing the dominant variations relative to the average (singular vector 1).
    These vectors reveal the formation of $\mathit{K}/\mathit{K'}$ supermodes upon SSB (red/blue color).}
    \label{fig:SVD}
\end{figure*}

An important question is whether the relative intensities between the $\mathit{K}/\mathit{K'}$ lasing outputs are always locked to each other in the two lasers, or whether the two lasers instead are subject to SSB independently. Our Fourier-space imaging setup does not disentangle which of the two lasing areas contributes which intensity to the Fourier-space output. However, we recall that the single-shot real-space analysis in Fig.~\ref{fig:single_shot} shows that high-contrast $\mathit{K}/\mathit{K'}$-mode lasing comes with distinct real-space intensity envelopes for single lasing shapes. Therefore, to answer our question, one can examine correlations in the real-space intensity patterns across the time series, which might reveal the various real-space mode patterns present in the system. We address this problem by employing singular value decomposition (SVD), which is a common method used to reduce dimensionality in correlated datasets \cite{sirovich_low-dimensional_1987, buijs_super-resolution_2020, rohrich_uncertainty_2020}. The first singular vector from SVD relates to the ensemble- or time-averaged intensity distribution, and the following $N_{\text{ims}}-1$ vectors quantify the dominant variations relative to that average. Singular values represent the weights with which the singular vectors contribute as basis vectors that span the data. Figure~\ref{fig:SVD} shows the first two singular vectors obtained from SVD analysis on a 100-long single-shot time series of two coupled lasers (in the SI, where we also study SVD on a single triangle). The first singular vector for D = 5~$\upmu$m is shown in Fig.~\ref{fig:SVD}(a). It reveals the time-averaged real-space envelope, and indicates that the pair of closely spaced lasers behave like a single undercoupled laser with high intensity at the edges of the enclosing parallelogram and low intensity in its center. When viewed at high resolution (see inset in Fig.~\ref{fig:SVD}(a)), one can discern a hexagonal fine structure at the periodicity of the lattice ($a=500$ nm), as reported earlier for the average behavior of $\mathit{K}/\mathit{K'}$-point lasing~\cite{de_gaay_fortman_spontaneous_2024}. The second singular vector (Fig.~\ref{fig:SVD}(d)) shows the most common variation relative to the average: In each triangular lasing shape, it shows two `downward-pointing' triangular envelopes in blue and two `upward-pointing' triangular envelopes in red, clearly corresponding to the real-space envelopes associated with $\mathit{K}/\mathit{K'}$-point symmetry breaking (see Figure~\ref{fig:single_shot}). Higher-order singular vectors (see SI) have much smaller singular values. We therefore conclude that the two lasing areas spaced by D = 5~$\upmu$m (Fig.~\ref{fig:coupled_triangles}(a,b) and Fig.~\ref{fig:SVD}(a,d)) are locked in their SSB behavior. 

Fig.~\ref{fig:SVD}(b) and (c) show the singular vectors of lasing areas with increased separation. As in the case of D = 5~$\upmu$m, also for D = 10~$\upmu$m there are just two significant singular vectors: The time-averaged intensity, and the dominant variation that encodes the real-space envelopes associated with  $\mathit{K}/\mathit{K'}$-mode SSB in relative intensity. If the two lasers would perform SSB individually, there would also be a third significant singular vector, in which the sign of the real space envelope in one of the lasers would be flipped. In Fig. S5, we report the first 6 singular vectors, and we find that for D = 5~$\upmu$m there is no such significant component, indicating that SSB in the two lasers still occurs in lockstep, and not independently. In contrast, for the D = 15~$\upmu$m case, we do find a significant singular vector of symmetry opposite to singular vector 2 (see Fig. S5). For instance, in the 100-shot time series at hand, it occurs 16 times that this contribution dominates over singular vector 2, meaning that the real-space envelope indicates  $\mathit{K}$-point lasing in one triangular area, but $\mathit{K'}$-point lasing in the other. We thus conclude from our SVD analysis that for $D \leq 10$~$\upmu$m, the two lasers are always locked in their $\mathit{K}/\mathit{K'}$-point SSB in relative intensity, but that this is no longer the case when D = 15~$\upmu$m. This loss of intensity synchronization is consistent with the reduced mutual coherence observed from the Fourier-space fringe pattern, which points at reduced phase synchronization. 

To summarize, our combined Fourier-/real-space approach to study coupled metasurface lasers demonstrates the capability to spatially control and image the nonlinear interaction dynamics of coupled lattice lasers. Therefore, our method unlocks additional degrees of freedom in exploration of such systems, going beyond previous works on complex nonlinear dynamics in coupled lasers \cite{mork_route_1990, heil_chaos_2001, happ_two-dimensional_2003, hamel_spontaneous_2015, ohtsubo_semiconductor_2017} --- studying, e.g., self-pulsations, chaos, synchronization, and SSB --- since in these works, the dynamics are typically only mapped via temporal studies of emission output.

\section{Theoretical model}\label{secDMD:model}

\begin{figure*}
    \centering
    \includegraphics[width=1\textwidth]{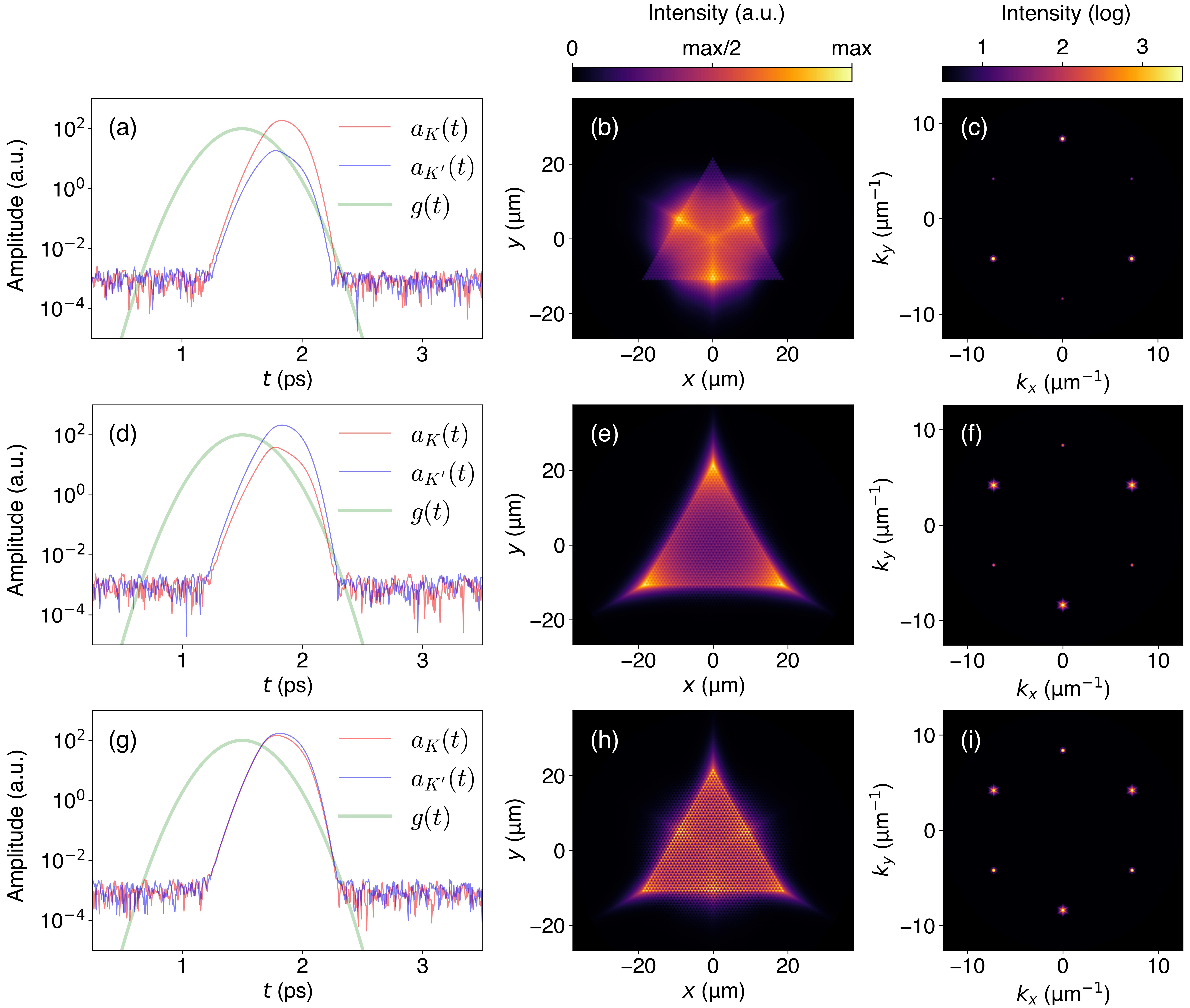}
    \caption{\textbf{Example numerical results obtained from the theoretical model.} Panels (a-c) correspond to single-shot lasing events with dominant $\mathit{K}$-point lasing, panels (d-f) to dominant $\mathit{K'}$-point lasing, and (g-i) to a balanced contribution of both. The first column (a,d,g) reveals the time evolution of the $a_{\mathit{K(K')}}$ mode amplitudes (red and blue curves, absolute values), together with the incident pump pulse $g(t)$ (green). The middle column (b,e,h) shows the time-averaged real-space intensity envelopes. The right column (c,f,i) shows the simulated far-field intensity, calculated as the Fourier transform of the instantaneous field at the peak of each lasing event. The far-field intensity is plotted in the logarithmic color scale. The Python code for the model is provided in the SI.}
    \label{fig:theoretical_results}
\end{figure*}

This section presents results obtained with a theoretical model that reproduces some key experimental observations presented in this paper, such as the real-space $\mathit{K}$- and $\mathit{K'}$-point lasing envelopes and the SSB in the relative intensity between $\mathit{K}$/$\mathit{K'}$ modes. We present all the details of the model in the SI. In brief, the model is an extension of the previous simple model introduced in~\cite{de_gaay_fortman_spontaneous_2024}, which captures SSB behavior for infinitely extended uniformly pumped $\mathit{K}$-point lattice laser by means of a density matrix approach in the tight-binding approximation, as first reported by Wu, Alpakov and Stockman~\cite{wu_topological_2020}. Here, we generalize the model to a lattice of finite dimensions with spatially varying pump, and include real-space propagation along $\mathit{K}$ (resp. $\mathit{K'}$) lattice directions. The propagation takes into account the instantaneous distribution of gain and loss across the lattice, depending on the local population inversion at a given instant of time. The pumping rate is defined individually for each lattice site, allowing us to consider arbitrary pump patterns.

Overall, the model has two assumptions derived from our empirical observations. The first assumption is that only the $\mathit{K}$- and $\mathit{K'}$-point Bloch modes compete for gain, so that the analysis can be limited to their spatially varying amplitude envelopes. This assumption is supported by the fact that there is no evidence in the data that any of the $\mathit{K}$-point lasers have other competing laser modes in the gain window (see SI). The second assumption is based on the observation in Fig.~\ref{fig:single_shot}, and stipulates that intensity builds up in the direction opposite to wave vector. This is equivalent to stating a negative group velocity as property of the Bloch modes near the photonic stop gap edge at which the lasing occurs. We implement our model in the form of a Python code, which is presented in the SI. The formalism entails a set of stochastic differential equations governing the evolution of the complex mode amplitudes, their intensity envelopes, as well as the population inversion and density matrix at each lattice site. We numerically solve the differential equations using the fixed-step Euler-Heun method within Python diffeqpy package~\cite{rackauckas2017differentialequations},  accelerated by numba Just-in-Time compilation~\cite{lam2015numba} (see SI). The lattice consists of $N$ = 20401 sites and forms a circular area of diameter equal to 150 lattice periods (all model parameters tabulated in SI, Table I). The system is seeded by random noise, uniformly applied to the entire circular simulation domain.  In the presented examples, the pumping rate distribution forms an equilateral triangle of base length 75 periods (37.5 $\upmu$m), comparable to the experimental configuration in Fig.~\ref{fig:single_shot}(a,b). The equations are solved with a time step $dt$ = 0.005 ps, which provides sufficient numerical stability. Each simulation of a single-shot lasing event takes around 10 minutes on a laptop.

Figure~\ref{fig:theoretical_results} shows three example realizations of the model, one giving rise to a dominant $\mathit{K}$-point lasing (a-c), another resulting in a dominant $\mathit{K'}$-point lasing (d-f), and a third random realization yielding an almost balanced contribution of both lasing modes (g-i). The real-space intensity distributions in Fig.~\ref{fig:theoretical_results}(b,e,h) are in excellent agreement with the experimental observations presented in Fig.~\ref{fig:single_shot}. The Fourier transforms of the modeled real-space distributions are shown in Fig.~\ref{fig:theoretical_results}(c,f,i) and show the concomitant $\mathit{K}/\mathit{K'}$-point lasing. As the model is scalar, the obtained Fourier-space images do not contain donut-shaped beams, but spots that are bright in their centers.

To verify if the SSB behavior of the modeled laser uniformly samples all linear combinations of $\mathit{K}$- and $\mathit{K'}$-point lasing, we have performed a Kolmogorov-Smirnov test on 100 numerical realizations of single-shot lasing. The results, presented in the SI, show that the intensity ratio ($\theta$) is random and uniformly distibuted over its full range, in line with the SSB behavior for infinite lattices observed in Ref.~\cite{de_gaay_fortman_spontaneous_2024}. On the other hand, the relative phase ($\phi$), which for infinite lasers is also uniformly distributed, tends to be concentrated around a fixed value, essentially pinned by the outer boundary of the pump area. This model allows us to verify the robustness of SSB against symmetry breaking by pump shape. Furthermore, by running our model for different parameters, one can easily obtain a variety of statistical behaviors, including strong biases and/or bifurcation, especially with smaller lasing areas. This demonstrates that our $\mathit{K}$-point lattice lasers are a unique experimental platform to realize, program, and study SSB in photonic lattices. The fact that the relative phase between $\mathit{K}$/$\mathit{K'}$ modes can be locked at a specific value, while in relative intensity the metasurface laser displays unbiased SSB, could have applications for on-chip optical simulators that combine stochastic ($\theta$) and deterministic ($\phi$) behavior. Our model can easily be adapted to explore diverse and exotic regimes of lasing dynamics, e.g., coupled lasers exhibiting nonlinear dynamics, or parity-time symmetry breaking, enabling the design of further experiments.

\section{Summary and conclusions}
Our work establishes spatially programmable plasmonic metasurface lasers as a versatile platform for exploring nonlinear and symmetry-driven phenomena in photonic lattices. By dynamically shaping the optical pump beam that delineates the gain area, we demonstrate control over lasing geometry, the vortex nature of the output beams, and spontaneous symmetry breaking (SSB) both in isolated and in coupled systems. The remarkable robustness of $\mathit{K}/\mathit{K'}$-point SSB, unbiased by pump geometry and resilient across diverse symmetry-lowering gain configurations, reveals a unique symmetry-protected degeneracy, tracing back to the fact that $\mathit{K}/\mathit{K'}$-point Bloch modes are not just spectrally but also spatially degenerate, as they are identical up to time-reversal. This robustness should be contrasted to the more generic case illustrated by $\mathit{M}$-point lasing, archetypical for most special points in photonic lattices. $\mathit{M}$-point lasing showcases deterministic control via spatially asymmetric pumping, owing to the lack of spatial degeneracy. Furthermore, the emergence of phase and amplitude synchronization between spatially separated K-point metasurface lasers highlights new avenues for studying coherence in coupled, extended, nonlinear systems, with exquisite control over the Bloch-mode mediated coupling between lasing areas, enabled by spatial programming of the pump beam.

These findings pave the way for several exciting directions. First, our platform offers a natural route to investigate SSB and the spontaneous emergence of coherence in non-Hermitian lattices, where gain and loss distributions can be precisely engineered. We envision that in addition to shaping laser boundaries, one can also address gain and loss per meta-atom, essentially projecting gain-loss superlattices over the sample at hand. Such studies could illuminate the interplay between topology, nonlinearity, and symmetry breaking in active photonic systems. The programmable nature of our approach invites extensions to lattices with alternative symmetries, including square, Lieb, honeycomb, kagome, or quasi-crystalline geometries, enabling exploration of topological lasing, exceptional points, and even synthetic gauge fields. These systems not only deepen our understanding of light-matter interaction in structured media, but also offer a rich playground for future photonic technologies that harness symmetry, topology, and nonlinearity.

In combination with our theoretical model, which captures the stochastic dynamics and spatial mode competition, this work lays the foundation for a new class of reconfigurable, nonlinear metasurface lasers. Beyond fundamental physics, metasurface lasers exhibiting SSB hold promise for on-chip applications in optical logic, neuromorphic computing, and true random number generation. The stochastic yet hard-to-bias nature of $\mathit{K}/\mathit{K'}$-point lasing offers a robust physical entropy source for random number generation, while the ability to synchronize or decouple lasing regions suggests potential for scalable photonic networks with programmable interactions.

\section{Acknowledgments}
This work is part of the Dutch Research Council (NWO) and was performed at the research institute AMOLF. We gratefully acknowledge G. Krause, B. Krijger, and M. Kamp for important contributions to the experimental setup. R.K. acknowledges support of the Research Council of Finland (Grants No. 347449, 353758 and 368485). The calculations were performed using computer resources within the Aalto University School of Science “Science-IT” project.

\bibliography{bibliography}

\end{document}


\title{Metasurface lasers programmed by optical pump patterns \\- Supporting Information}

\author{Nelson de Gaay Fortman}
    \affiliation{Institute of Physics, University of Amsterdam, 1098 XH Amsterdam, The Netherlands} 
    \affiliation{Department of Physics of Information in Matter and Center for Nanophotonics, NWO-I Institute AMOLF, Science Park 104, NL1098XG Amsterdam, The Netherlands}
   \author{Radoslaw Kolkowski}
    \affiliation{Department of Applied Physics, Aalto University, P.O. Box 13500, FI-00076 Aalto, Finland}
\author{Nick Feldman}
    \affiliation{Department of Physics of Information in Matter and Center for Nanophotonics, NWO-I Institute AMOLF, Science Park 104, NL1098XG Amsterdam, The Netherlands}
    \affiliation{Advanced Research Center for Nanolithography (ARCNL), Science Park 106, 1098 XG Amsterdam, The Netherlands}
\author{Peter Schall}
    \affiliation{Institute of Physics, University of Amsterdam, 1098 XH Amsterdam, The Netherlands}
\author{A. Femius Koenderink\footnote[1]{Corresponding author: f.koenderink@amolf.nl} }
    \affiliation{Department of Physics of Information in Matter and Center for Nanophotonics, NWO-I Institute AMOLF, Science Park 104, NL1098XG Amsterdam, The Netherlands}
    \affiliation{Institute of Physics, University of Amsterdam, 1098 XH Amsterdam, The Netherlands}

\date{\today}
\maketitle
 
\tableofcontents
\newpage

\begin{figure*}
    \centering
    \includegraphics[width=1\textwidth]{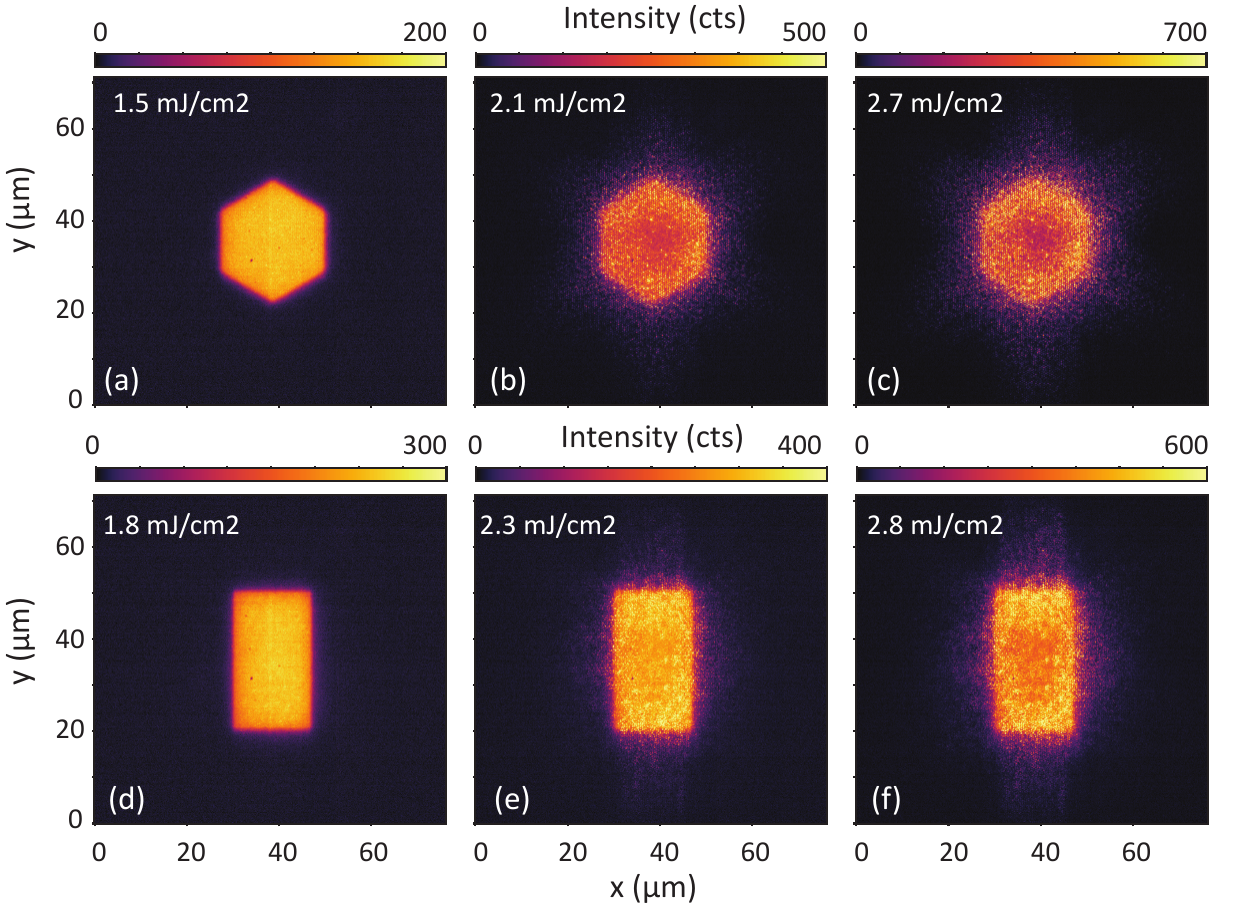}
    \caption{\textbf{Evolution of real-space intensity distributions with pump fluence.} (a-c) Hexagons of short diagonal 23 $\upmu$m. (e-f) Rectangles of shape ($17.3\times30$) $\upmu$m. At the lowest fluence the signal is due to fluorescence, not lasing. At the intermediate fluence both fluorescence and lasing contribute comparably.}
    \label{SIfig:fluence_evolution}
\end{figure*}
 
\subsection{Measurement protocol}\label{secDMD:protocol}
To obtain the results presented in Figs.~2--5 of the main article, we performed lasing experiments for various pump area shapes and sizes following a specific protocol. Before exploring any new pump 
pattern, we navigated to a fresh, previously unpumped patch in the 200$\times$200 $\upmu$m$^2$ metasurface area, to avoid the effects of photobleaching from previous experiments. When measuring a pattern of a given shape but for a series of sizes, we started with measurement for the smallest size and then performed the measurements sequentially for monotonically increasing sizes in the same area. When investigating pairs of coupled triangular lasing areas, we first performed the measurements for the shortest separation distance, and then we did the measurements for the increasing  separations while staying centered on the same area.

\subsection{Real-space distributions as function of pump fluence}\label{secDMD:envelope_fluences}
In coupled wave theory for DFB lasers, the amount of gain influences the nature of the lasing modes (undercoupled, critically coupled, overcoupled). What is more, at higher fluences, various mode envelope solutions may exist above threshold and compete for the gain. In Fig.~\rk{S}\ref{SIfig:fluence_evolution}, we study the evolution of the real space intensity envelopes with increasing pump fluences. For both  hexagons (Fig.~S\ref{SIfig:fluence_evolution}a-c) and rectangles (Fig.~S\ref{SIfig:fluence_evolution}d-f), we observe undercoupled envelopes that do not drastically change their profile for higher gain.  This may point at other mechanisms that shape the real-space envelopes beyond standard 1D Kogelnik and Shank theory~\cite{guo_spatial_2019}, attributable to the nontrivial Bloch wave properties at the K-points.
 
\begin{figure*}
    \centering
    \includegraphics[width=1\textwidth]{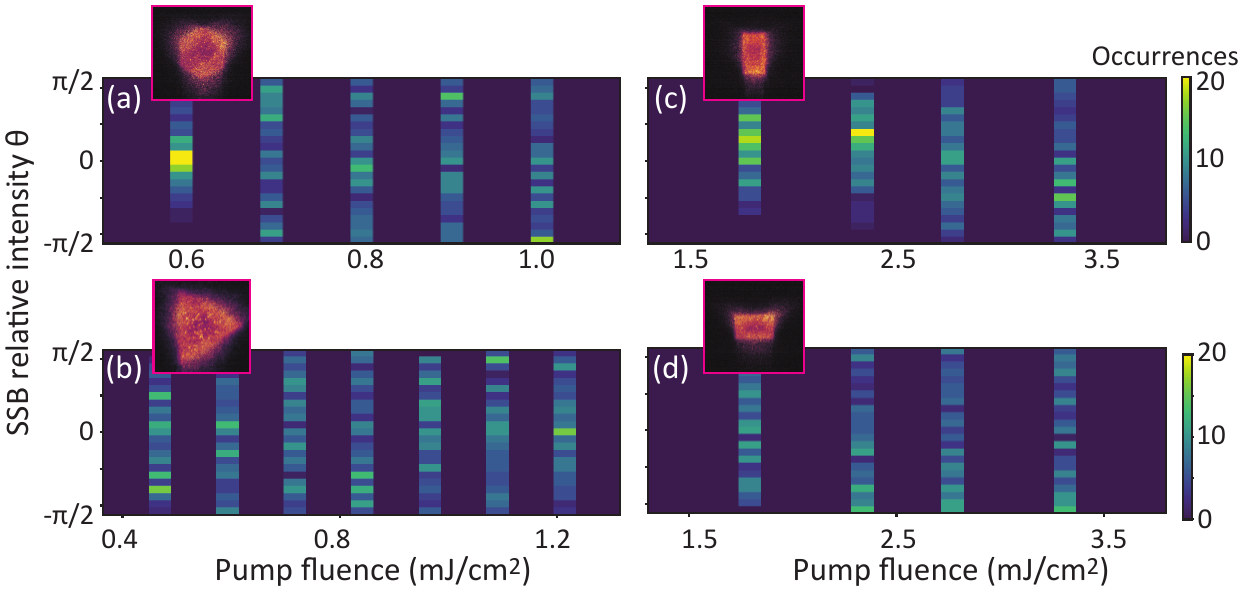}
    \caption{\textbf{Statistics of SSB  in relative  intensity.} Single-shot time traces of the SSB parameter $\theta$ that traces the relative 
     intensity between $\mathit{K}$ and $\mathit{K'}$, for the (a)  individual   hexagon (short diagonal 34.5 $\upmu$m), (b) rotated triangle (side length 53 $\upmu$m), and (c,d) vertical and horizontal  rectangle ($17.3\times30$) $\upmu$m.}
    \label{SIfig:SSB_statistics}
\end{figure*}

\subsection{SSB statistics}\label{secDMD:SSB_stats}
In Fig.~S\ref{SIfig:SSB_statistics}, we display histograms for the parity SSB statistics for single-shot time traces belonging to pump patterns presented in Fig.~3 of the main article. This SSB is expressed in  terms of parameter $\theta$ that maps the relative intensity of $\mathit{K}$- and $\mathit{K'}$-point lasing intensity onto the interval $-\pi/2<\theta<\pi/2$. We plot $\theta$ as a function of pump fluence for $\geq100$ shots per investigated fluence. Figure~S\ref{SIfig:SSB_statistics} shows the results for hexagon (Figure~S\ref{SIfig:SSB_statistics}a), right-oriented triangle (b), upward triangle (c) and downward triangle (d). None of these patterns exhibit a significant tendency towards either $\mathit{K}$- or $\mathit{K'}$-point lasing. 
 
\subsection{Relative phase fit for coupled lasers}\label{secDMD:phi_R_coupled}
\begin{figure*}
    \centering
    \includegraphics[width=1\textwidth]{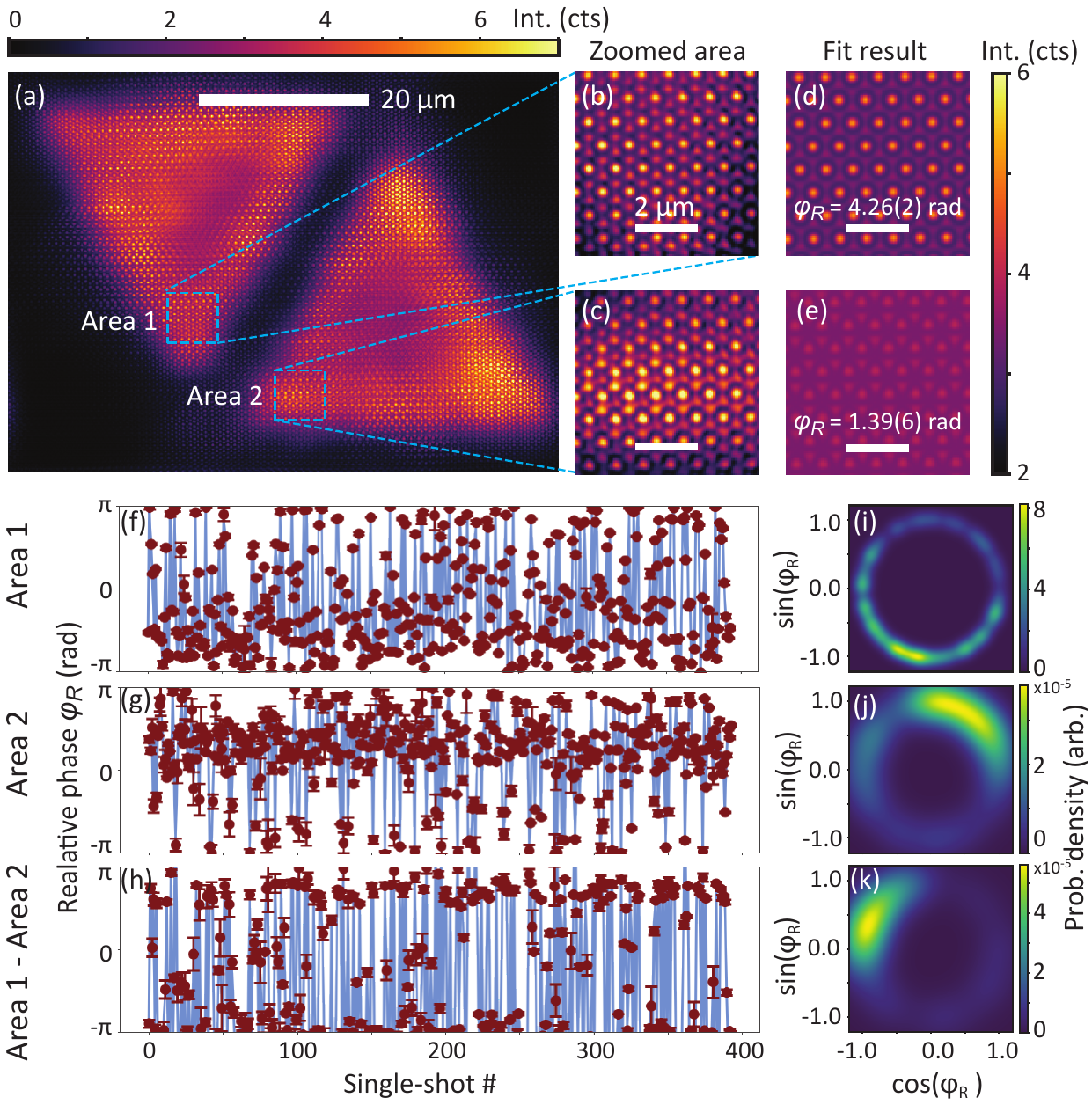}
    \caption{\textbf{Relative phase between $\mathit{K}$ and $\mathit{K'}$ measured on the border of each triangle.} (a) Fourier-filtered real-space image, same data as main Fig.~5, D = 10 $\upmu$m. (b,c) Enlargements of two blue areas in (a). (d,e) Fit results to extract relative phase $\varphi_\text{R}$. (f,g) 400-long time series of fitted $\varphi_\text{R}$ from area 1 and 2, hinting at phase bias to values near $-2\pi/3$ and $\pi/3$, respectively. (h) Difference in measured $\varphi_\text{R}$, biasing towards $-\pi$. (i-k) Probability density plots of $\varphi_\text{R}$ as projected on the unit circle.}
    \label{SIfig:relative_phase_fit}
\end{figure*} 

For the two L = 35 $\upmu$m triangular lasers with separation distances D = 10 $\upmu$m and D = 15 $\upmu$m, clear fringes appear in the Fourier-space donut-spots (main Fig.~5). We attribute the appearance of these fringes to mutual coherence between the lasers. Since the fringe pattern does not significantly vary within the recorded time trace of single-shot lasing events, we argue that the two lasers synchronize with a definite, reproducible phase. We further substantiate this claim by measuring the relative phase between $\mathit{K}$ and $\mathit{K'}$ on a border area of both triangles by analysis of real-space images, using the protocol laid out in detail in the supplement of Ref. \cite{de_gaay_fortman_spontaneous_2024}. 

First, we Fourier-filter the real-space image at hand and find that the Fourier transform displays 25 components. We mask the Fourier components with a Gaussian filter of width 1.07 $\upmu$m$^{-1}$, and then we inverse-Fourier transform the image back to the real space (see single-shot example in Fig.~S\ref{SIfig:relative_phase_fit}). The Gaussian width corresponds to a real-space resolution of 5.86 $\upmu$m, or 100 pixels. In Fig.~S\ref{SIfig:relative_phase_fit}(b,c) we show close-ups of 5.86 $\upmu$m square areas selected in Fig.~S\ref{SIfig:relative_phase_fit}(a). Here, the microscopic detail of the filtered real-space image reveals superlattice patterns that we recognize from Ref. \cite{de_gaay_fortman_spontaneous_2024} as interference between the plane-wave triplets of $\mathit{K}$ and $\mathit{K'}$. In Ref. \cite{de_gaay_fortman_spontaneous_2024}, we have shown that by fitting the plane waves of the $\mathit{K}$ and $\mathit{K'}$ modes, $$E_{\text{total}}(\mathbf{r}) = |a_{\mathit{K}}| E_{\mathit{K}}(\mathbf{r}) + |a_{\mathit{K'}}| E_{\mathit{K'}}(\mathbf{r})e^{i\varphi_\text{R}},$$
to the hexagon/honeycomb superlattice patterns, one can retrieve the relative phase $\varphi_\text{R}$. The results are shown in Fig.~S\ref{SIfig:relative_phase_fit}(d,e). We apply this method to a time-trace of 400 single shots, extracting $\varphi_\text{R}$ in both regions. Already in this time series, a clear bias in $\varphi_\text{R}$ is visible for both areas: Area 1 tends towards $-2\pi/3$ (Fig.~S\ref{SIfig:relative_phase_fit}(f)), whereas area 2 towards $\pi/3$ (Fig.~S\ref{SIfig:relative_phase_fit}(g)). Their difference appears to be fixed at around $-\pi$ (Fig.~S\ref{SIfig:relative_phase_fit}(h)). The phase bias in the time series is better visualized by projecting the phase on the unit circle \cite{de_gaay_fortman_spontaneous_2024}. In these projections, shown in Fig.~\rk{S}\ref{SIfig:relative_phase_fit}(i-k), the phase fit error sets the width of the unit circle, and the resulting histogram displays the probability density of measured phases, clearly reflecting the biased behavior. This analysis demonstrates that, whereas in infinite systems both the $\mathit{K}/\mathit{K'}$-mode relative intensity and phase are random and unbiased upon SSB, for these finite, coupled systems the geometry leaves the $\mathit{K}/\mathit{K'}$-intensity unbiased, but can pin the difference between the two measured phases, which indicates phase locking.

\begin{figure*}
    \centering
    \includegraphics[width=1\textwidth]{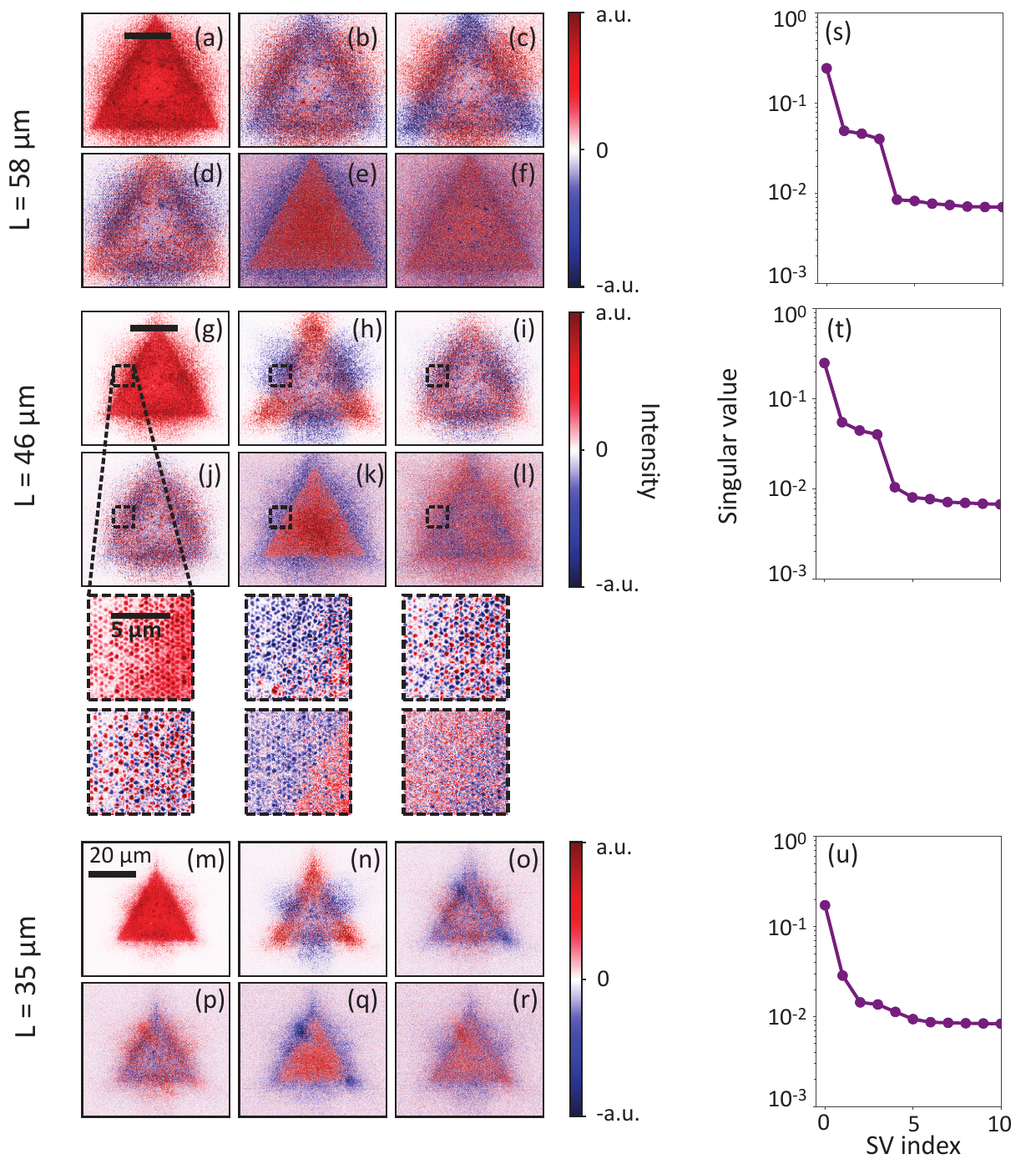}
    \caption{\textbf{Singular value decomposition (SVD)  results for 
    individual triangular lasers.} (a-f) First six singular vectors  obtained from SVD performed  on a single-shot real-space lasing image of 
     an individual triangle of L = 58 $\upmu$m. (g-l) First six singular vectors for a 46 $\upmu$m triangle, together with six enlarged images in which hexagonal features on a microscopic level are visible. (m-r) Singular value decomposition results for a  L = 35 $\upmu$m. (s-u) First ten singular values associated to vectors displayed on the left.}
    \label{SIfig:SVD_single_laser}
\end{figure*}

\subsection{Singular value decomposition (SVD) of real-space intensity distributions}\label{secDMD:SVD}
We apply the SVD  method to the real-space lasing images in order to find similarities in SSB behavior among the single shots within a given time trace. We first study individual triangularly shaped lasers (see Fig.~S\ref{SIfig:SVD_single_laser}) of side lengths 
L = 58 $\upmu$m (first six singular vectors shown in  Fig.~S\ref{SIfig:SVD_single_laser}(a-f)), 
L = 46 $\upmu$m (Fig.~S\ref{SIfig:SVD_single_laser}(g-l)), and
L = 35 $\upmu$m (Fig.~S\ref{SIfig:SVD_single_laser}(m-r)). 
For a time series of 100 shots, we find that the first six singular vectors are responsible for almost all of the signal in the lasing emission and fluorescence background, as indicated by the singular value plots in Fig.~S\ref{SIfig:SVD_single_laser}(s-u). Singular vectors beyond the 6$^{\text{th}}$ essentially correspond to the noise floor of the  single-shot data, capturing just random (shot) noise of fluorescence and lasing speckle. From the singular vector plots in Fig.~\rk{S}\ref{SIfig:SVD_single_laser}, it can be seen that the boundaries and corners become increasingly sharp as the triangle area gets smaller.  On the whole, the SVD clearly indicates that the dominant variation relative to the mean intensity pattern lies in the three-lobed envelope associated to the $\mathit{K}/\mathit{K'}$-mode lasing, indicating that the real-space envelope blinks in sync with the wave vector output, as discussed in the main paper. In close-up images (shown for L = 46 $\upmu$m, Fig.~S\ref{SIfig:SVD_single_laser}(g-i) insets), we discern hexagonal intensity patterns on the unit-cell level with high resolution. Vectors 3 and 4 (Fig.~S\ref{SIfig:SVD_single_laser}(i,j)) show hexagonal features in both negative (blue) and positive (red) intensities. These reconstitute the hexagonal/honeycomb superlattice patterns that we also observe after Fourier filtering (see Fig.~S\ref{SIfig:relative_phase_fit}).

\begin{figure*}
    \centering
    \includegraphics[width=1\textwidth]{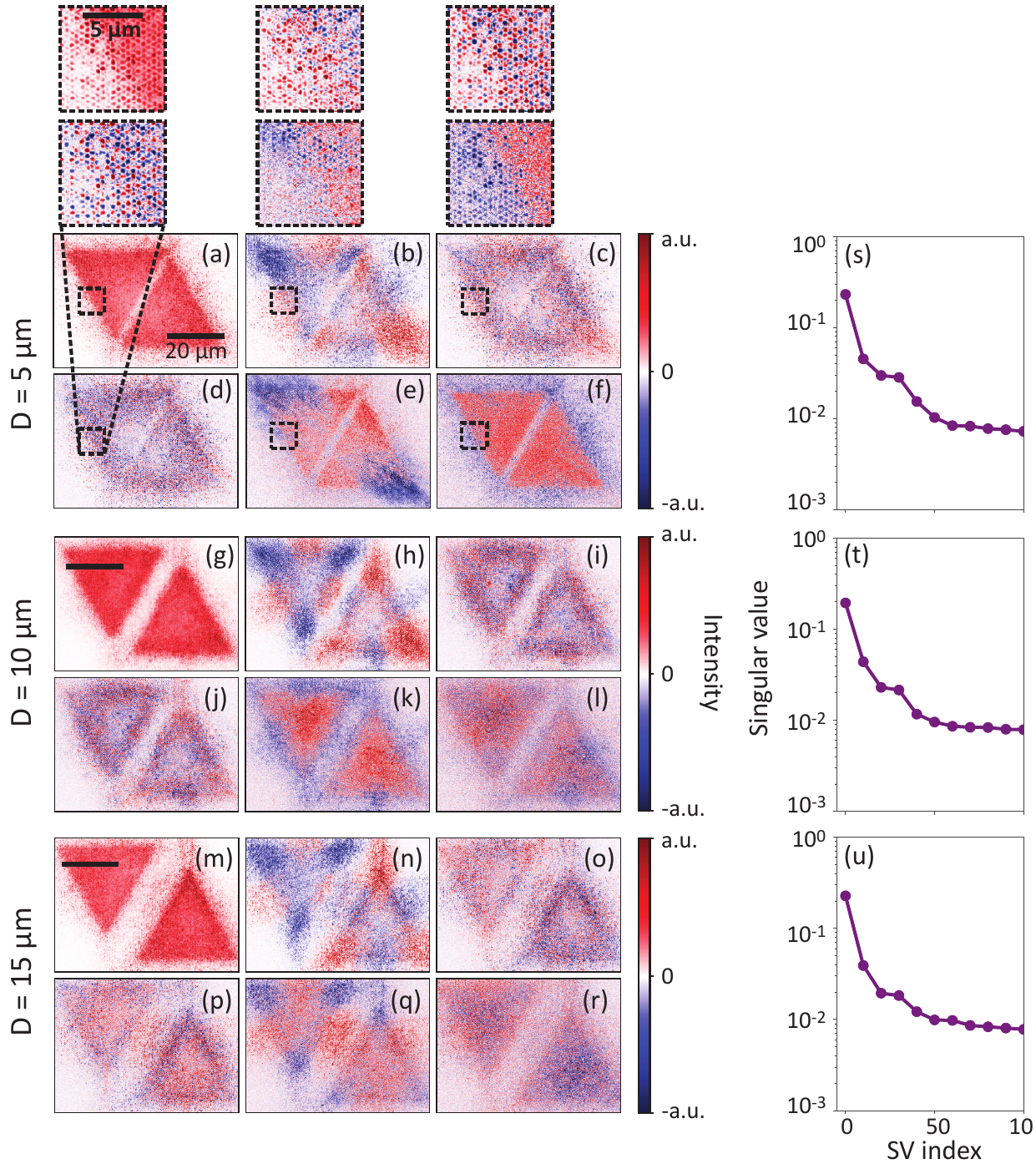}
    \caption{\textbf{Overview of SVD results for coupled triangular lasers,} with side length L = 35 $\upmu$m and increasing separation distance. (a-f) First six singular vectors for D = 5 $\upmu$m, with enlarged areas that display microscopic detail. (g-l) Singular vectors for D = 10 $\upmu$m. (m-r) Singular vectors for D = 15 $\upmu$m. (s-u) Singular values for the three separation distances.}
    \label{SIfig:SVD_two_lasers}
\end{figure*}

Finally, in Fig.~S\ref{SIfig:SVD_two_lasers}, we provide 
an  overview of the singular vectors and values corresponding to the coupled triangular lasers shown in main manuscript Fig.~5. We perform SVD on 100-long single-shot sequences. 
Figure~S\ref{SIfig:SVD_two_lasers}(a-f) shows  singular vectors for separation distance D = 5 $\upmu$m, together with insets highlighting microscopic, hexagonal details. Figure~S\ref{SIfig:SVD_two_lasers}(a) represents the ensemble average of the sequence, and shows an intensity dip in the center of the total lasing area, which is a signature of undercoupled lasing. In contrast, when D = 10 $\upmu$m or 15 $\upmu$m, Fig.~S\ref{SIfig:SVD_two_lasers}(g) and (m) show the same signature in both lasing areas   individually. Importantly, for D = 5 $\upmu$m and D = 10 $\upmu$m, only one singular vector contributes to the typical triangular intensity distribution associated with $\mathit{K}/\mathit{K'}$-mode lasing, namely 
Fig.~S\ref{SIfig:SVD_two_lasers}(b) and (h) respectively. This is no longer the case when D = 15 $\upmu$m, since Fig.~S\ref{SIfig:SVD_two_lasers}(q) shows a pattern that allows $\mathit{K}$-point lasing in one triangle and at the same time $\mathit{K'}$-point lasing in the other.

\subsection{Theoretical model}\label{secDMD:model}

The model assumes the lasing dynamics to be governed by the following set of stochastic differential equations:
\begin{widetext}
\begin{equation}\label{eq:1}
d a_{\mathit{K(K')}}(t)=-\gamma\, a_{\mathit{K(K')}}(t)dt+i\,\Omega \frac{N_{g}}{N}\sum_{l}^{N}\sqrt{e_{l,\mathit{K(K')}}(t)}\left[\bar{\rho}_l(t)\psi_{l,\mathit{K(K')}}\right]^*dt+\sigma\, dW_{t,\mathit{K(K')}}
\end{equation}
\begin{equation}\label{eq:2}
\frac{d e_{l,\mathit{K(K')}}(t)}{dt}=\zeta\operatorname{prop}_{l,\mathit{K(K')}}[\mathbf{e}_{\mathit{K(K')}}(t),\mathbf{n}(t)]+\xi\,\frac{1+n_l(t)}{|a_{\mathit{K}}|^2+|a_{\mathit{K'}}|^2+\sigma^2}+\eta\left\{1-\frac{1}{2N}\sum_l^N[e_{l,\mathit{K}}(t)+e_{l,\mathit{K'}}(t)]\right\}
\end{equation}
\begin{equation}\label{eq:3}
\frac{dn_l(t)}{dt}=g_l(t)[1-n_l(t)]-\gamma_{\text{nr}}[1+n_l(t)]-4\Omega\operatorname{Im}\left\{\bar{\rho}_l(t)\left[\sqrt{e_{l,\mathit{K}}(t)}a_{\mathit{K}}(t)\psi_{l,\mathit{K}}+\sqrt{e_{l,\mathit{K'}}(t)}a_{\mathit{K'}}(t)\psi_{l,\mathit{K'}}\right]\right\}
\end{equation}
\begin{equation}\label{eq:4}
\frac{d\bar{\rho}_l(t)}{dt}=-\Gamma_{12}\bar{\rho}_l(t)+i\, n_l(t)\Omega\left[\sqrt{e_{l,\mathit{K}}(t)}a_{\mathit{K}}(t)\psi_{l,\mathit{K}}+\sqrt{e_{l,\mathit{K'}}(t)}a_{\mathit{K'}}(t)\psi_{l,\mathit{K'}}\right]^*
\end{equation}
\end{widetext}
Similarly to the previous works~\cite{wu_topological_2020,de_gaay_fortman_spontaneous_2024}, the gain medium is treated here as a two-level system described by a spatially-varying density matrix $\rho_l(t)$ in the rotating wave approximation, in which $n_l(t)=\rho_{22,l}(t)-\rho_{11,l}(t)$ is the population inversion at a lattice site $l$, $\bar{\rho}_l(t)$ is the complex amplitude of the non-diagonal density matrix element $\rho_{12,l}(t)$, $\gamma$ is the decay rate of the $\mathit{K}$/$\mathit{K'}$ mode amplitudes $a_{\mathit{K(K')}}(t)$ (inversely proportional to the mode $Q$ factor), $\Omega$ is the Rabi frequency, quantifying the rate of coupling between the gain medium and the modes (depending the product of the mode's electric field and the transition dipole moment of the gain medium molecules), $N_g$ is the number of the gain medium molecules contributing to the amplification of $a_{\mathit{K(K')}}(t)$, $g_l(t)$ is the pumping rate at a lattice site $l$ with time dependence corresponding to a Gaussian pulse $g_l(t) = g_l^{(0)}g_{\text{max}}\exp[-(t-t_0)^2/\tau^2]$, with a binary mask $g_l^{(0)}\in\{0,1\}$ defining the pump pattern, $\gamma_{\text{nr}}$ is the nonradiative decay rate, and $\Gamma_{12}$ is the polarization relaxation rate.

Equation~(\ref{eq:1}) describes the temporal evolution of complex amplitudes $a_{\mathit{K}}(t)$ and $a_{\mathit{K'}}(t)$. The wavefunctions of the corresponding modes are defined as the product of their instantaneous real-space amplitude envelopes $\sqrt{e_{l,\mathit{K(K')}}(t)}$ (normalized by setting $\sum_l^N e_{l,\mathit{K(K')}}(t)/N=1$) and the infinite-lattice wavefunctions $\psi_{\mathit{K(K')}}$:
\begin{equation}
\psi_{l,K}=\psi_{l,K'}^*=\frac{1}{3}\left(e^{i\mathbf{K}_1\cdot\mathbf{r}_l}+e^{i\mathbf{K}_2\cdot\mathbf{r}_l}+e^{i\mathbf{K}_3\cdot\mathbf{r}_l} \right)
\end{equation}
where $\mathbf{K}_1=\frac{4\pi}{3\Lambda}(1,0)$, $\mathbf{K}_{2,3}=\frac{2\pi}{3\Lambda}(-1,\pm\sqrt{3})$, $\Lambda$ = 500 nm is the lattice period, $\mathbf{r}_l$ are the real-space coordinates of the lattice sites, $\mathbf{r}_l=p_l\mathbf{u}+q_l\mathbf{v}$, with $\mathbf{u}=\Lambda(1,0)$ and $\mathbf{v}=\frac{\Lambda}{2}(1,\sqrt{3})$ being the lattice vectors and $p_l$, $q_l$ spanning over a range of integers. The last term in Eq.~(\ref{eq:1}), $\sigma\, dW_{t,\mathit{K(K')}}$, represents a complex-valued additive noise of amplitude $\sigma$, resulting from two uncorrelated Wiener processes $dW_{t,\mathit{K}}$ and $dW_{t,\mathit{K'}}$.

While Equations~(\ref{eq:3}) and (\ref{eq:4}) represent the usual rate equations governing the dynamics of population inversion and density matrix in a laser medium (see Refs.~\cite{wu_topological_2020,de_gaay_fortman_spontaneous_2024}), Equation~(\ref{eq:2}) constitutes a nonstandard addition to the model, addressing the dynamics of normalized real-space intensity envelopes $\mathbf{e}_{\mathit{K(K')}}(t)$. The first term of Eq.~(\ref{eq:2}) represents a redistribution of intensity across the lattice due to the propagation of $\mathit{K}$ and $\mathit{K'}$ Bloch waves. The rate of this redistribution is determined by the parameter $\zeta$. We define a $\mathit{K}$/$\mathit{K'}$ propagation operator, ``$\operatorname{prop}$'', which modifies the local intensity $e_l$ at a lattice site $l$ by accumulating the intensities of lattice sites $l'$ located along straight lines intersecting at $l$ and parallel to the $\mathit{K}$/$\mathit{K'}$ directions. The algorithm takes into account net attenuation/amplification of the intensities on their ways from $l'$ to $l$, based on the instantaneous real-space distribution of the population inversion $\mathbf{n}(t)$. The action of the ``$\operatorname{prop}$'' operator can be written in two steps:
\begin{equation}
e_{l,\mathit{K(K')}}(t+\Delta t)=\sum_{l'[l,K(K')]}e_{l',\mathit{K(K')}}(t)\exp\left\{\sum_{l''(l,l')}G[n_{l''}(t)]\right\}
\end{equation}
\begin{equation}\label{eq:7}
\operatorname{prop}_{l,\mathit{K(K')}}\left[\mathbf{e}_{\mathit{K(K')}}(t),\mathbf{n}(t)\right]=\frac{\sum_l^N e_{l,\mathit{K(K')}}(t)}{\sum_l^N e_{l,\mathit{K(K')}}(t+\Delta t)}e_{l,\mathit{K(K')}}(t+\Delta t)
\end{equation}
The first step represents the accumulation of intensities from lattice sites $l'$ (selection of which depends on $l$ and $\mathit{K}$/$\mathit{K'}$), each contributing with a net gain and/or loss $G[n_{l''}(t)]$ summed over the lattice sites $l''$ located in between $l$ and $l'$. Furthermore, the ``$\operatorname{prop}$'' operator is constructed in a way that avoids edge effects resulting from finite extent of the lattice (see the Python code implementation in the further sections below). 
The gain/loss is assumed to depend on the population inversion as follows:
\begin{equation}
G[n_{l''}(t)]=\frac{\,n_{l''}(t)+|n_{l''}(t)|}{2}G_{\text{prop}}+\frac{\,n_{l''}(t)-|n_{l''}(t)|}{2}L_{\text{prop}}
\end{equation}
with amplification coefficient $G_{\text{prop}}$ contributing when $n_{l''}(t)>0$ and attenuation coefficient $L_{\text{prop}}$ contributing when $n_{l''}(t)<0$. In the second step (Eq.~(\ref{eq:7})), the envelope is renormalized to maintain a constant total intensity. Apart from the propagation term, Equation~(\ref{eq:2}) contains two auxiliary terms, one responsible for seeding by spontaneous emission (proportional to $1+n_l(t)$), and the other enforcing renormalization of both envelopes ($\mathit{K}$ and $\mathit{K'}$) during simulations. The contributions of these two terms are set by parameters $\xi$ and $\eta$, respectively.

\begin{table}[th!]
\caption{\label{tab:t1} Parameter values used for the example calculations to obtain the results presented in Fig.~7 and Fig.~S\ref{fig:theoretical_stats}.}
\begin{center}
\begin{tabular}{|lcc|}
\,\,Parameter\,\,\,\, &\,\,\,\,Value\,\,\,\,&\,\,\,\,Unit\,\,\,\,\\
\colrule
\,\,$\Omega$ & 0.1 & ps$^{-1}$ \\ 
\,\,$\gamma$ & 2 & ps$^{-1}$ \\ 
\,\,$N_g$ & 10$^6$ & 1 \\ 
\,\,$N$ & 20401 & 1\\ 
\,\,$\sigma$ & 0.01 & ps$^{-1}$\\ 
\,\,$dt$ & 0.005 & ps\\ 
\,\,$\zeta$ & 3 & ps$^{-1}$\\ 
\,\,$\xi$ & 0.01 & ps$^{-1}$\\
\,\,$\eta$ & 200 & ps$^{-1}$\\ 
\,\,$\gamma_{\text{nr}}$ & 5 & ps$^{-1}$ \\ 
\,\,$\Gamma_{12}$ & 200 & ps$^{-1}$ \\
\,\,$t_0$ & 1.5 & ps \\ 
\,\,$\tau$ & 0.25 & ps \\
\,\,$g_{\text{max}}$ & 100 & ps$^{-1}$ \\
\,\,$G_{\text{prop}}$ & 0.075 & 1 \\ 
\,\,$L_{\text{prop}}$ & 0.15 & 1 \\ 
\,\,$a_{l,\mathit{K(K')}}(0)$ & 0 & 1 \\ 
\,\,$e_{l,\mathit{K(K')}}(0)$ & 1 & 1 \\ 
\,\,$n_{l}(0)$ & $-$1 & 1 \\ 
\,\,$\bar{\rho}_{l}(0)$ & 0 & 1 \\ 
\end{tabular}
\end{center}
\end{table}

\subsection{Kolmogorov-Smirnov test of the theoretical results}\label{secDMD:code}

\begin{figure*}
    \centering
    \includegraphics[width=0.8\textwidth]{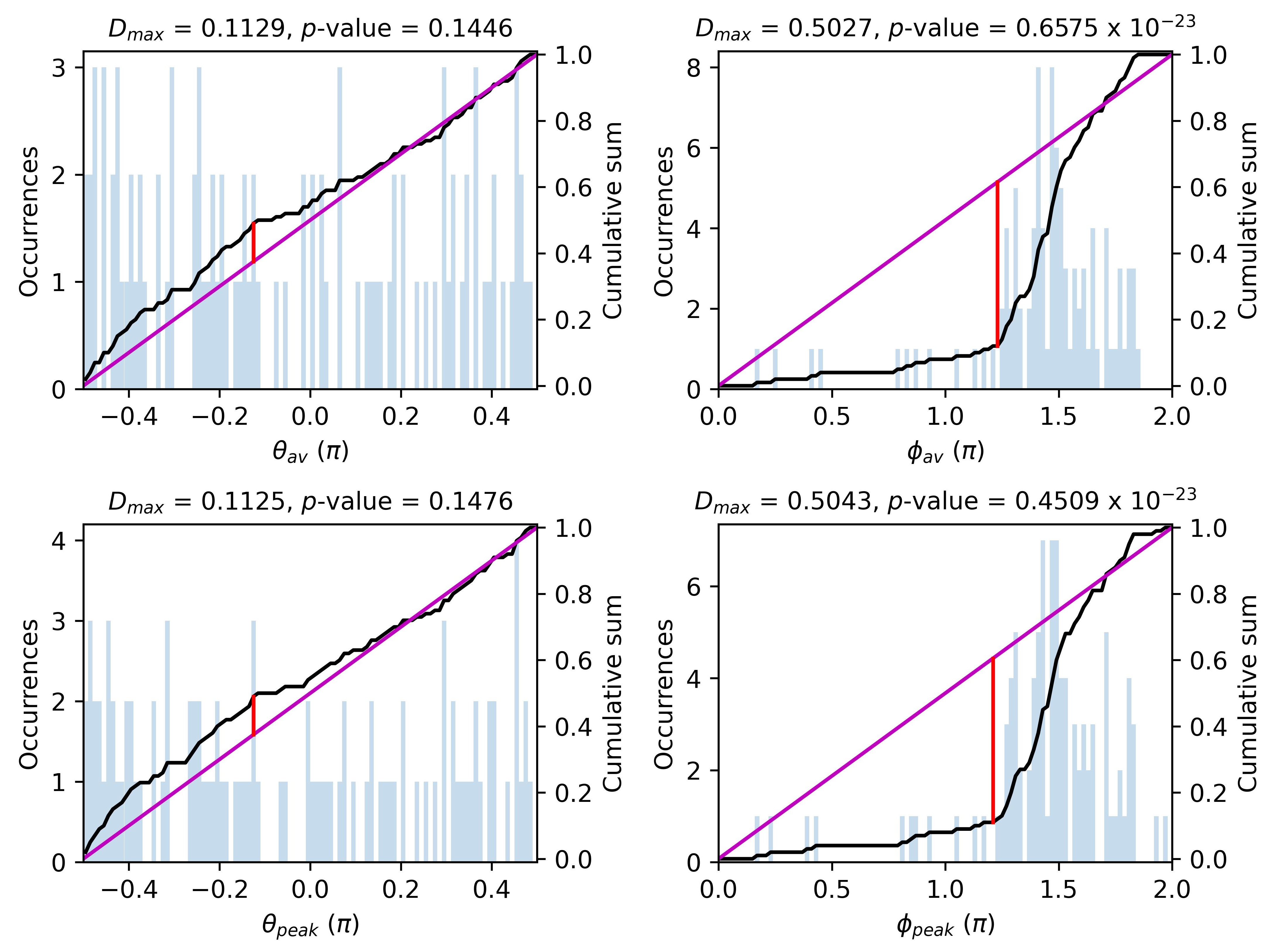}
    \caption{\textbf{Statistical analysis of 100 numerical realizations of single-shot lasing using our theoretical model} for the same system as in Fig.~7 with parameters specified in Table~\ref{tab:t1}. The analysis is performed using one-sample Kolmogorov-Smirnov (KS) test against the null hypothesis of a uniform distribution of $\theta$ and $\phi$ (which quantify the relative $\mathit{K}$/$\mathit{K'}$ intensity and phase, respectively). We consider both time-averaged and peak values of $a_{\mathit{K}(\mathit{K'})}$, yielding $\theta_{\text{av}}$, $\phi_{\text{av}}$ and $\theta_{\text{peak}}$, $\phi_{\text{peak}}$, respectively. The KS test shows that both $\theta_{\text{av}}$ and $\theta_{\text{peak}}$ are uniformly distributed ($p$-values $>$ 0.05), as opposed to $\phi_{\text{av}}$ and $\phi_{\text{peak}}$ ($p$-values close to zero).}
    \label{fig:theoretical_stats}
\end{figure*}

Figure~S\ref{fig:theoretical_stats} presents statistical analysis of 100 realizations of the model, confirming the SSB behavior of this lasing system. The analysis is carried out using a one-sample Kolgomorov-Smirnov (KS) test. The obtained KS p-values are greater than the 0.05 significance level, which indicates that the relative $\mathit{K}$/$\mathit{K'}$ intensities, quantified by the $\theta$ parameter, can be regarded as uniformly distributed over their full range (from $-\pi/2$ to $+\pi/2$), although a subtle bias towards the $\mathit{K'}$-mode can be seen from the cumulative sum graphs. This statistical behavior is observed in $\theta$ calculated both from the peak and time-averaged intensities ($\theta_{\text{peak}}$, $\theta_{\text{av}}$). On the other hand, the relative phase $\phi$ between the $\mathit{K}$ and $\mathit{K'}$ modes appears to be significantly clustered around a specific value, which is consistent with the experimental observation that, compared to infinite systems where SSB in both intensity and phase is unbiased (see Ref.~\cite{de_gaay_fortman_spontaneous_2024}), constrained pump geometries leave the intensity SSB unbiased but can cause pinning of the phase values (see Fig.~S\ref{SIfig:relative_phase_fit} in this supplement, displaying the results for triangle pairs).

\subsection{Python code for the theoretical model}\label{secDMD:code}

Here we present our Python code, which implements the theoretical model presented in the previous section, allowing one to simulate the $\mathit{K}$-point lasing dynamics with a finite gain pattern defined over a uniform hexagonal lattice. The code is distributed into four separate files:
\begin{enumerate}
\item\vspace{-2mm} \lstinline[basicstyle=\normalsize\ttfamily]{'K_lasing_model.py'} - the main script that defines the lattice and gain distribution, reads parameter values and settings from the \lstinline[basicstyle=\normalsize\ttfamily]{'parameters.py'} file, models the lasing dynamics by solving stochastic differential equations, and plots/saves the results;
\item\vspace{-2mm} \lstinline[basicstyle=\normalsize\ttfamily]{'lattice_analysis.py'} - script containing function \lstinline[basicstyle=\normalsize\ttfamily]{find_sites_along_K}, which is used by the main script to find lattice sites that contribute with their local field intensity to the propagation along $\mathit{K}$ and $\mathit{K'}$ directions towards each site in the lattice;
\item\vspace{-2mm} \lstinline[basicstyle=\normalsize\ttfamily]{'propagation_algorithm.py'} - script containing function \lstinline[basicstyle=\normalsize\ttfamily]{prop}, which modifies an input intensity distribution by propagating it along $\mathit{K}$ and $\mathit{K'}$ directions based on the collections of lattice site indices obtained from \lstinline[basicstyle=\normalsize\ttfamily]{find_sites_along_K};
\item\vspace{-2mm}  \lstinline[basicstyle=\normalsize\ttfamily]{'parameters.py'} - file containing most of the model parameters and some settings for plotting/saving the results, read by the main script via function \lstinline[basicstyle=\normalsize\ttfamily]{get_params}.

\end{enumerate}

\begin{figure*}
    \centering
    \includegraphics[width=0.8\textwidth]{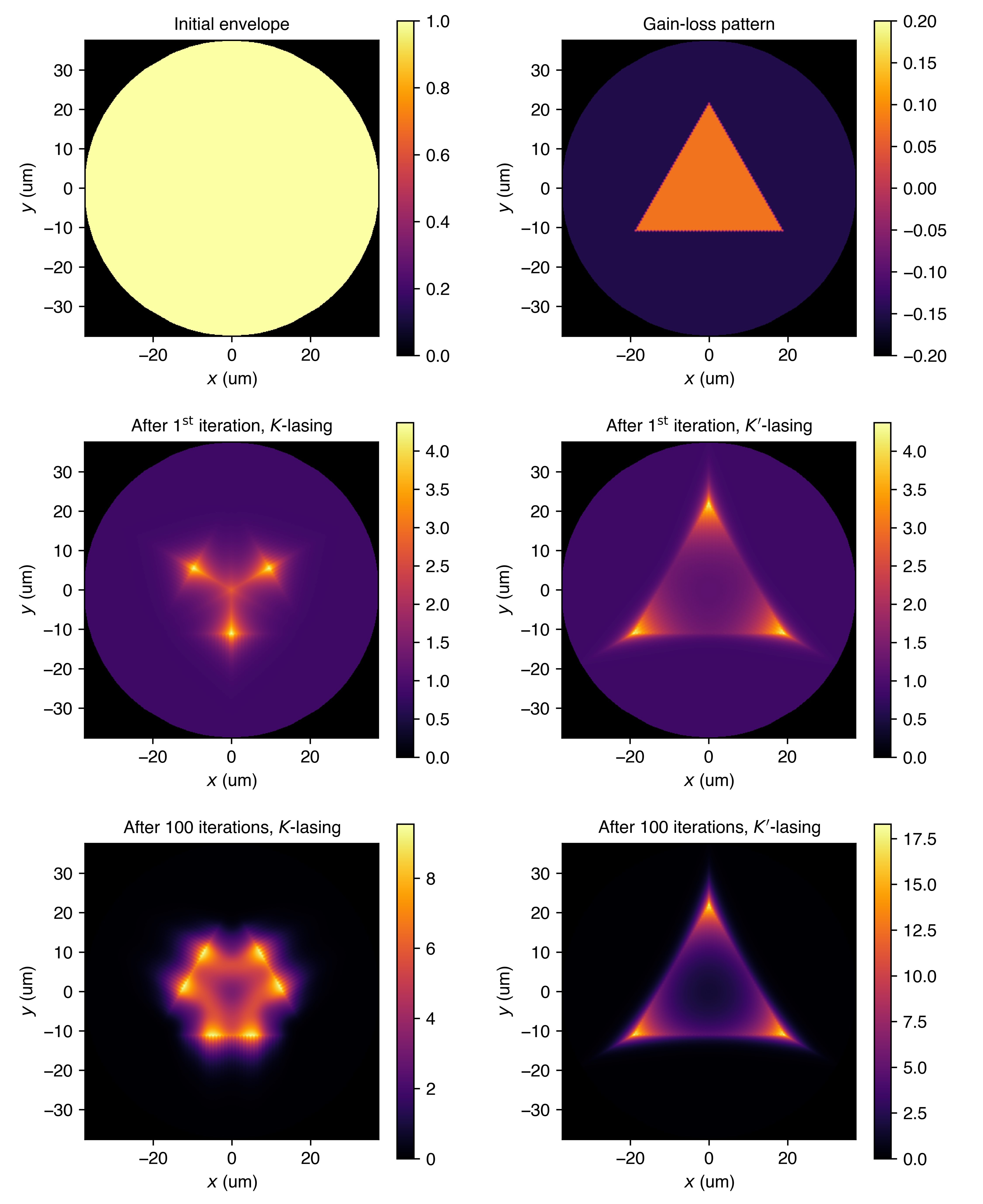}
    \caption{\textbf{Illustration of the propagation algorithm.} The initial envelope $\mathbf{e}$ for both the $\mathit{K}$ and $\mathit{K'}$ modes is defined as a uniform distribution across the lattice, whereas the gain/loss pattern $\mathbf{g}$ is shaped into a triangle (with base length equal to 75 lattice periods). In each iteration, a new envelope $\mathbf{e'}_{\mathit{K(K')}}$ is obtained from the previous envelope $\mathbf{e}_{\mathit{K(K')}}$ through the following operation: $\mathbf{e'}_{\mathit{K(K')}}=(1-p)\mathbf{e}_{\mathit{K(K')}}+p\operatorname{prop}_{\mathit{K(K')}}
    [\mathbf{e}_{\mathit{K(K')}},\mathbf{g}\,]$, with $p$ set to 0.2. In the complete simulation of lasing dynamics using the Python code provided below, the gain distribution $\mathbf{g}$ evolves together with population inversion $\mathbf{n}(t)$.}
    \label{SIfig:prop_action}
\end{figure*}

To run the code without errors, all necessary packages must be installed. These include the \lstinline[basicstyle=\normalsize\ttfamily]{diffeqpy} package, which is a Python interface to \lstinline[basicstyle=\normalsize\ttfamily]{DifferentialEquations.jl} -- an advanced Julia suite for solving differential equations~\cite{rackauckas2017differentialequations}. Another important package used by the code is \lstinline[basicstyle=\normalsize\ttfamily]{numba}~\cite{lam2015numba}, which significantly accelerates the \lstinline[basicstyle=\normalsize\ttfamily]{prop} function, making simulations feasible. The lasing dynamics are simulated by running \lstinline[basicstyle=\normalsize\ttfamily]{'K_lasing_model.py'} after defining the lattice and gain pattern in the first few lines. At this point, the gain and loss settings are defined to illustrate the action of the \lstinline[basicstyle=\normalsize\ttfamily]{prop} function in the first figure generated by the code -- see example in Fig.~S\ref{SIfig:prop_action}. These gain/loss settings can be modified based on the results displayed in the figure, and then implemented by saving and rerunning the main script. Other settings and parameter values are set in the \lstinline[basicstyle=\normalsize\ttfamily]{'parameters.py'} file. By default, the units for space, time, and frequency are $\upmu$m, ps, and ps$^{-1}$ (THz), respectively. Since importing \lstinline[basicstyle=\normalsize\ttfamily]{diffeqpy} may take from a few seconds up to a few minutes (due to activation of Julia environment), the essential part of the code (solving differential equations, plotting/saving the results) is placed inside a loop that starts after \lstinline[basicstyle=\normalsize\ttfamily]{diffeqpy} is imported. At the beginning of each repetition of this loop, the model settings and parameters are pulled from the \lstinline[basicstyle=\normalsize\ttfamily]{'parameters.py'} file, which can be updated and saved while the code is paused during displaying the results (if "\lstinline[basicstyle=\normalsize\ttfamily]{show = 1}"). There is also another, inner loop which repeats the simulation a number of times (defined by "\lstinline[basicstyle=\normalsize\ttfamily]{rep}") for a fixed set of parameters, which can be used, e.g., to explore the $\mathit{K}$/$\mathit{K'}$-lasing SSB statistics. An example code to perform Kolgomorov-Smirnov test on $\mathit{K}$/$\mathit{K'}$ relative intensities ("\lstinline[basicstyle=\normalsize\ttfamily]{sv = 1}") is provided at the end of this section (an additional code \lstinline[basicstyle=\normalsize\ttfamily]{'read_amplitudes.py'}).

\vspace{10mm}

Contents of the Python files:

\begin{enumerate}
\item\lstinline[basicstyle=\normalsize\ttfamily]{'K_lasing_model.py'}
\begin{lstlisting}
import matplotlib.pyplot as p
from numpy import array as a
from numpy import linspace as ls
from numpy import meshgrid as mg
from numpy import exp,conj,pi,real,imag,log10,dot,int32,zeros,angle
from scipy.interpolate import LinearNDInterpolator as ND
from time import time

P=0.5 # Lattice period
Nxy=75 # Lattice extent (approx. radius, # of periods) > 0
Nx,Ny=Nxy,Nxy # Lattice x and y extent
Nmax=max(Nx,Ny)
x,y=ls(-2*Nmax*P,2*Nmax*P,num=4*Nmax+1),\
     ls(-2*Nmax*P,2*Nmax*P,num=4*Nmax+1)
X,Y=mg(x,y)
rX,rY=X+Y/2,Y*3**(1/2)/2
Npix = 500 # Resolution of the interpolated plots, # of pixels

# Arrays of x and y lattice coordinates, circular
rx=rX[(abs(rX+1j*rY)<=Nmax*P)].ravel()
ry=rY[(abs(rX+1j*rY)<=Nmax*P)].ravel()

# Gain distribution (pump pattern)
b=Nx*P # Triangle size (base length)
gcoeff = 0.075 # Gain coefficient inside the pumped area
lcoeff = 0.150 # Loss coefficient elsewhere
g=rx*0.0-lcoeff
g[(rx>=-b*3**(1/2)/6)*(rx<=(ry+b/3)*3**(1/2))*\
  (rx<=-(ry-b/3)*3**(1/2))]=gcoeff

plot_prop = 1 # Illustration of the propagation algorithm: 0 - no, 1 - yes
iterations = 100 # Number of iterations for the illustration, > 2
prdt = 0.2 # Propagation rate x time step (for the illustration)

# K vectors
K1=4*pi/3/P*a([1,0])
K2=2*pi/3/P*a([-1,3**(1/2)])
K3=2*pi/3/P*a([-1,-3**(1/2)])

# K and K' wavefunctions
def psi(xy):
    return (exp(1j*(K1[0]*xy[0]+K1[1]*xy[1]))+\
            exp(1j*(K2[0]*xy[0]+K2[1]*xy[1]))+\
            exp(1j*(K3[0]*xy[0]+K3[1]*xy[1])))/3

psi1=a(list(map(lambda xy: psi(xy), list(zip(rx,ry)))))
psi2=conj(psi1)


# Lattice analysis

N = len(rx)
m = int32(max([(max(rx)-min(rx))/P,(max(ry)-min(ry))/P])+1)

print("\nFinding sets of lattice site indices for K/K' propagation:\n"+\
      "\n* number of lattice sites = "+str(N)+\
      "\n* max. number of indices in each set = "+str(m)+\
      "\n* max. total number of indices in all sets = "+\
      str(6)+" x "+str(N)+" x "+str(m)+" = "+str(6*N*m)+ "\n...")

t0=time()
from lattice_analysis import find_sites_along_K
p1,p2,p3,p4,p5,p6=find_sites_along_K(rx,ry,m)
t1=time()

np1,np2,np3 = len(p1[(p1!=-1)]),len(p2[(p2!=-1)]),len(p2[(p2!=-1)])
np4,np5,np6 = len(p4[(p4!=-1)]),len(p5[(p5!=-1)]),len(p6[(p6!=-1)])
nptot = np1+np2+np3+np4+np5+np6

print('... completed in ' +'%.6f'%(t1-t0)+' sec\n'+\
      '\n* total number of collected indices in all sets = '+str(nptot)+\
      '\n  using '+'%.2f'%(nptot/(6*N*m)*100)+\
      ' % of allocated array space\n')


# Precompilation of the propagation algorithm

from propagation_algorithm import prop


print("\nPrecompilation of the propagation algorithm\n...")

F00=rx*0.0+1.0
t0=time()
F1=prop(F00,g,p4,p5,p6,m)
t1=time()
print('... completed in ' +'%.6f'%(t1-t0)+' sec\n')



F00=rx*0.0+1.0
t0=time()
for i in range(iterations//2):
    if i==0:
        F01=F00*1.0
        F02=F00*1.0
    else:
        F01=F1*1.0
        F02=F2*1.0
    if i==1:
        F11=F1*1.0
        F12=F2*1.0
    elif i==2:
        F21=F1*1.0
        F22=F2*1.0
    F1=F01*(1-prdt)+prdt*prop(F01,g,p4,p5,p6,m)
    F2=F02*(1-prdt)+prdt*prop(F02,g,p1,p2,p3,m)
t1=time()

print('* average of '+str(iterations)+' subsequent runs = '+\
          '%.6f'%((t1-t0)/(iterations-1))+' sec\n')


Rx,Ry=ls(min(rx),max(rx),num=Npix*Nx//Nmax),\
       ls(min(ry),max(ry),num=Npix*Ny//Nmax)
RX,RY=mg(Rx,Ry) # Dense grid for interpolation (used only for plotting)

Flist=[F00,g,F11,F12,F21,F22,F1,F2]
ptlist=['Initial envelope','Gain-loss pattern',\
        r'After 1$^{\text{st}}$ iteration, $K$-lasing',\
        r"After 1$^{\text{st}}$ iteration, $K'$-lasing",\
        r'After 2$^{\text{nd}}$ iteration, $K$-lasing',\
        r"After 2$^{\text{nd}}$ iteration, $K'$-lasing",\
        r'After '+str(iterations)+' iterations, $K$-lasing',\
        r"After "+str(iterations)+" iterations, $K'$-lasing"]
        
        

if plot_prop == 1:

    fig0=p.figure(figsize=(16,8))
    p.suptitle('Illustration of the propagation algorithm, '+\
               'assuming intensity build-up in opposite direction, '+\
               r"i.e., along $K$ in $K'$-lasing, along $K'$ in $K$-lasing",\
               fontsize=10)

    for i in range(len(Flist)):

        p.subplot(2,4,i+1)
        Fi=ND(list(zip(rx,ry)),Flist[i])
        F_=Fi(RX,RY)
        p.pcolormesh(RY,RX,F_,cmap='inferno') # Flipping x and y axes
        p.colorbar()
        if i!=1:
            p.clim(0,max(Flist[i]))
        else:
            p.clim(-4/3*max(lcoeff,gcoeff),4/3*max(lcoeff,gcoeff))
        p.xlabel(r'$x$ (um)')
        p.ylabel(r'$y$ (um)')
        p.gca().set_aspect('equal')
        p.gca().set_facecolor('k')
        p.title(ptlist[i],fontsize=10)

    p.subplots_adjust(left=0.05,right=0.95,wspace=0.3,hspace=0.3)
    fig0.savefig('prop_illustration.png',format='png',dpi=600)
    p.show()

#'''
# Lasing dynamics

def pl(x): # Function to apply to the plotted mode amplitudes
    return log10(x)

print('\nActivating diffeqpy (may take some time, please be patient)\n...')
time0=time()
from diffeqpy import de
time1=time()
print('... completed in '+'%.6f'%(time1-time0)+' sec')

solver = de.EulerHeun()

from numpy import absolute as ab
import shelve
import os
from importlib import reload
import parameters
from datetime import datetime

keep_repeating = True

while keep_repeating == True:

    parameters = reload(parameters)
    params_ = parameters.get_params(rx,ry,b,g)
    Om,Ng,gK,gnr,G12,pr,sr,rr,t0,tau,gmax,dt,noise,a0,b0,\
    tspan,F0,gp,lp,G,sv,pm,rep,show,pt = params_

    G12_ = G*0.0 + G12

    sumF0 = sum(F0)

    # Pumping rate as a function of time
    def gt(t): 
        return t*0+gmax*(ab(exp(-(t-t0)**2/tau**2)\
                -exp(-25))+(exp(-(t-t0)**2/tau**2)-exp(-25)))/2
    
    # Initial values for the differential equations
    p0 = [a0,b0]+[-1.0]*N+[0.0+0.0j]*N+F0.tolist()+F0.tolist()

    # Differential equations of the lasing system
    def f(dy,y,pms,t):
        
        dy[0]=-gK*y[0]+1.0j*Ng*Om*conj(dot(psi1*ab(y[2*N+2:3*N+2])**(1/2),y[N+2:2*N+2]))/N
        
        dy[1]=-gK*y[1]+1.0j*Ng*Om*conj(dot(psi2*ab(y[3*N+2:4*N+2])**(1/2),y[N+2:2*N+2]))/N
        
        dy[2:N+2]=G*gt(t)*(1.0-real(y[2:N+2]))-gnr*(1.0+real(y[2:N+2]))-\
                     4*Om*imag(y[N+2:2*N+2]*(y[0]*psi1*ab(y[2*N+2:3*N+2])**(1/2)+\
                                             y[1]*psi2*ab(y[3*N+2:4*N+2])**(1/2)))

        dy[N+2:2*N+2]=-G12_*y[N+2:2*N+2]+\
                     +1.0j*real(y[2:N+2])*Om*conj(y[0]*psi1*ab(y[2*N+2:3*N+2])**(1/2)+\
                                                y[1]*psi2*ab(y[3*N+2:4*N+2])**(1/2))

        
        dy[2*N+2:3*N+2]=pr*prop(ab(y[2*N+2:3*N+2]),\
                                gp*(real(y[2:N+2])+ab(real(y[2:N+2])))/2.0\
                            +lp*(real(y[2:N+2])-ab(real(y[2:N+2])))/2.0,p4,p5,p6,m)\
                            +sr*(1.0+real(y[2:N+2]))/(ab(y[0])**2+ab(y[1])**2+noise**2)\
                            +rr*(1.0-sum(ab(y[2*N+2:4*N+2]))/sumF0/2)*y[2*N+2:3*N+2]
        
        dy[3*N+2:4*N+2]=pr*prop(ab(y[3*N+2:4*N+2]),\
                                gp*(real(y[2:N+2])+ab(real(y[2:N+2])))/2.0\
                            +lp*(real(y[2:N+2])-ab(real(y[2:N+2])))/2.0,p1,p2,p3,m)\
                            +sr*(1.0+real(y[2:N+2]))/(ab(y[0])**2+ab(y[1])**2+noise**2)\
                            +rr*(1.0-sum(ab(y[2*N+2:4*N+2]))/sumF0/2)*y[3*N+2:4*N+2]
    
    # Wiener noise
    def W(dy,y,pms,t):
        dy[0,0] = noise
        dy[1,1] = noise
    
    for rpt in range(rep):
        
        print('\n\nSolving differential equations\n...')
        time0=time()
        prob = de.SDEProblem(f, W, p0, tspan, [],
                             noise_rate_prototype=(0.0+0.0j)*zeros((len(p0),2)))
        sol = de.solve(prob, solver, dt=dt)
        time1=time()
        print('... completed in '+'%.6f'%(time1-time0)+' sec')

        us=a(de.transpose(de.stack(sol.u)))

        
        Nt = len(sol.t)

        if pm==1:

            fig=p.figure(figsize=(10,8))
            plot1=p.subplot(111)

        elif pm==2:

            fig=p.figure(figsize=(16,8))
            plot1=p.subplot(121)

        if pm==1 or pm==2:

            p.suptitle(pt,fontsize=10)

            env_sum1 = a([sum(ab(us[tt,2*N+2:3*N+2]))/sumF0 for tt in range(Nt)])
            env_sum2 = a([sum(ab(us[tt,3*N+2:4*N+2]))/sumF0 for tt in range(Nt)])
                
            ax0,=p.plot(sol.t,pl(gt(a(sol.t))),'g',alpha=0.25,lw=2)
            ax1,=p.plot(sol.t,pl(ab(us[:,0])),'r',lw=1,alpha=0.5)
            ax2,=p.plot(sol.t,pl(ab(us[:,1])),'b',lw=1,alpha=0.5)
            ax10,=p.plot(sol.t,env_sum1,':r',lw=1,alpha=0.5)
            ax20,=p.plot(sol.t,env_sum2,':b',lw=1,alpha=0.5)
            
            for i in range(N):
                p.plot(sol.t,real(us[:,i+2]),'orange',lw=1,alpha=0.01)

            ax3,=p.plot([-1],[0],'orange',lw=1)
            ax4,=p.plot(sol.t,(angle(us[:,0])-\
                               angle(us[:,1]))%(2*pi)/pi,'--k',lw=1,alpha=0.5)

            p.legend([ax0,ax1,ax2,ax10,ax20,ax3,ax4],['Pumping rate: '+\
                                        r'$\operatorname{log}[g(t)]$',\
                                        'Mode amplitude: '+\
                                        r'$\operatorname{log}|a_{K}|$',\
                                        'Mode amplitude: '+\
                                        r"$\operatorname{log}|a_{K'}|$",\
                                        r'$K$ envelope sum',\
                                        r"$K'$ envelope sum",\
                                        'Population inversion: $n$',\
                                        r"$K$/$K'$ relative phase ($\pi$)"],\
                     framealpha=0,ncol=3,loc='upper left')

            p.xlabel('Time (ps)')
            p.ylabel('Solution')
            p.ylim(-4,4)
            p.xlim(min(sol.t),max(sol.t))

        if pm==2:

            Ntt = 0

            F1av,F2av,F12av,nav = F0*0.0,F0*0.0,F0*0.0,F0*0.0
            for tt in range(Nt):
                Ntt+=1
                F1av += ab(ab(us[tt,2*N+2:3*N+2])**(1/2)*us[tt,0]*psi1)**2
                F2av += ab(ab(us[tt,3*N+2:4*N+2])**(1/2)*us[tt,1]*psi2)**2
                #F1av += ab(us[tt,2*N+2:3*N+2])
                #F2av += ab(us[tt,3*N+2:4*N+2])
                F12av += ab(ab(us[tt,2*N+2:3*N+2])**(1/2)*us[tt,0]*psi1+\
                            ab(us[tt,3*N+2:4*N+2])**(1/2)*us[tt,1]*psi2)**2
                nav += real(us[tt,2:N+2])

            Flist=[F1av/Ntt,F2av/Ntt,F12av/Ntt,nav/Ntt]
            ilist=[3,4,7,8]
            ptlist=[r'$K$ mode',\
                    r"$K'$ mode",\
                    r"$K$ + $K'$ mode",\
                    r'Population inversion']
                    
            for i in range(len(Flist)):
                p.subplot(2,4,ilist[i])
                Fi=ND(list(zip(rx,ry)),Flist[i])
                F_=Fi(RX,RY)
                p.pcolormesh(RY,RX,F_,cmap='inferno') # # Flipping x and y axes
                p.colorbar()
                p.xlabel(r'$x$ (um)')
                p.ylabel(r'$y$ (um)')
                p.gca().set_aspect('equal')
                p.gca().set_facecolor('k')
                p.title(ptlist[i],fontsize=10)


            p.subplots_adjust(left=0.05,right=0.95,wspace=0.3,hspace=0.3)

        if sv==2:

            database_dir_name = 'solution_'+str(rpt)+\
                                '_'+str(datetime.now().year)\
                                +'_'+str(datetime.now().month)+\
                                '_'+str(datetime.now().day)\
                                +'_'+str(datetime.now().hour)+\
                                '_'+str(datetime.now().minute)\
                                +'_'+str(datetime.now().second)
            os.mkdir(database_dir_name)
            current_dir = os.path.dirname(os.path.abspath(__file__))
            database_dir = os.path.join(current_dir, database_dir_name)
            database_name = 'solution.db'
            database_path = os.path.join(database_dir, database_name)
            database = shelve.open(database_path)
            database['t'] = sol.t
            database['aK1'] = us[:,0]
            database['aK2'] = us[:,1]
            database['n'] = us[:,2:2+N]
            database['rho'] = us[:,2+N:2+2*N]
            database['e1'] = us[:,2+2*N:2+3*N]
            database['e2'] = us[:,2+3*N:2+4*N]
            database['p'] = params_
            database['r'] = rx,ry
            database.close()

            if pm==1 or pm==2:

                fig_name = 'lasing_dynamics.jpg'
                fig.savefig(os.path.join(database_dir, fig_name),\
                            format='jpg',dpi=300)
                if show==0:
                    p.close(fig)
                else:
                    p.show()

        elif sv==1:

            file = open('amplitudes_'+str(rpt)+'.txt','a')
            for i in range(len(sol.t)):
                file.write('%.8f'%(sol.t[i])+'\t'+\
                           '%.8f'%(real(us[i,0]))+'\t'+\
                           '%.8f'%(imag(us[i,0]))+'\t'+\
                           '%.8f'%(real(us[i,1]))+'\t'+\
                           '%.8f'%(imag(us[i,1]))+'\n')
            file.close()

            

        if show==0 and rep>1 and rpt==rep-1:
            keep_repeating = False
        elif show==1 and rep==1:
            p.show()
    
\end{lstlisting}

\item\lstinline[basicstyle=\normalsize\ttfamily]{'lattice_analysis.py'}:
\begin{lstlisting}
from numpy import angle, pi, int32, argsort, zeros
from numpy import array as a

def find_sites_along_K(rx,ry,m):
    
    M=zeros(m).astype(int32)
    MN=zeros(m*len(rx)).astype(int32)
    aI=a(range(len(rx)))
    P=abs(rx[1]-rx[0]+1j*(ry[1]-ry[0]))
    K1,K2,K3=4*pi/3/P*a([1,0]),2*pi/3/P*a([-1,3**(1/2)]),2*pi/3/P*a([-1,-3**(1/2)])
    a1,a2,a3,a4,a5,a6=angle(K1[0]+1j*K1[1]),angle(K2[0]+1j*K2[1]),angle(K3[0]+1j*K3[1]),\
                       angle(-K1[0]-1j*K1[1]),angle(-K2[0]-1j*K2[1]),angle(-K3[0]-1j*K3[1])
                       
    p1,p2,p3,p4,p5,p6=MN*0,MN*0,MN*0,MN*0,MN*0,MN*0
    
    for i in range(len(rx)):

        ai = angle(rx[i]-rx+1j*(ry[i]-ry))
        
        mask0 = (abs(rx[i]-rx+1j*(ry[i]-ry)) > P*1e-6)
        mask1 = (abs(ai-a1) < 1e-6)
        mask2 = (abs(ai-a2) < 1e-6)
        mask3 = (abs(ai-a3) < 1e-6)
        mask4 = (abs(ai-a4) < 1e-6)
        mask5 = (abs(ai-a5) < 1e-6)
        mask6 = (abs(ai-a6) < 1e-6)
        
        p1_unsorted = aI[mask0*mask1]
        p2_unsorted = aI[mask0*mask2]
        p3_unsorted = aI[mask0*mask3]
        p4_unsorted = aI[mask0*mask4]
        p5_unsorted = aI[mask0*mask5]
        p6_unsorted = aI[mask0*mask6]
        
        d1_unsorted = abs(rx[i]-rx[p1_unsorted]+1j*(ry[i]-ry[p1_unsorted]))
        d2_unsorted = abs(rx[i]-rx[p2_unsorted]+1j*(ry[i]-ry[p2_unsorted]))
        d3_unsorted = abs(rx[i]-rx[p3_unsorted]+1j*(ry[i]-ry[p3_unsorted]))
        d4_unsorted = abs(rx[i]-rx[p4_unsorted]+1j*(ry[i]-ry[p4_unsorted]))
        d5_unsorted = abs(rx[i]-rx[p5_unsorted]+1j*(ry[i]-ry[p5_unsorted]))
        d6_unsorted = abs(rx[i]-rx[p6_unsorted]+1j*(ry[i]-ry[p6_unsorted]))
        
        p1_sorted = p1_unsorted[argsort(d1_unsorted)]
        p2_sorted = p2_unsorted[argsort(d2_unsorted)]
        p3_sorted = p3_unsorted[argsort(d3_unsorted)]
        p4_sorted = p4_unsorted[argsort(d4_unsorted)]
        p5_sorted = p5_unsorted[argsort(d5_unsorted)]
        p6_sorted = p6_unsorted[argsort(d6_unsorted)]
        
        p1_,p2_,p3_,p4_,p5_,p6_ = M*0-1, M*0-1, M*0-1, M*0-1, M*0-1, M*0-1
        
        p1_[:len(p1_sorted)] = p1_sorted
        p2_[:len(p2_sorted)] = p2_sorted
        p3_[:len(p3_sorted)] = p3_sorted
        p4_[:len(p4_sorted)] = p4_sorted
        p5_[:len(p5_sorted)] = p5_sorted
        p6_[:len(p6_sorted)] = p6_sorted
        
        p1[i*m:(i+1)*m]=p1_
        p2[i*m:(i+1)*m]=p2_
        p3[i*m:(i+1)*m]=p3_
        p4[i*m:(i+1)*m]=p4_
        p5[i*m:(i+1)*m]=p5_
        p6[i*m:(i+1)*m]=p6_
        
    return p1,p2,p3,p4,p5,p6
\end{lstlisting}

\item\lstinline[basicstyle=\normalsize\ttfamily]{'propagation_algorithm.py'}:
\begin{lstlisting}
import numba
from numpy import exp, zeros, argmin

@numba.njit(parallel=True,fastmath=True)
def prop(F,G,p1,p2,p3,m):

    lenF=len(F)
    F1=zeros(lenF)
    
    for i in numba.prange(lenF):
        s1, s2, s3 = 0.0, 0.0, 0.0

        p1im = p1[i*m:(i+1)*m]
        p2im = p2[i*m:(i+1)*m]
        p3im = p3[i*m:(i+1)*m]

        j1 = argmin(p1im)-1
        j2 = argmin(p2im)-1
        j3 = argmin(p3im)-1

        pj1 = p1im[j1]
        pj2 = p2im[j2]
        pj3 = p3im[j3]

        f_1 = F[pj1]
        f_2 = F[pj2]
        f_3 = F[pj3]
        
        g_1 = G[pj1]
        g_2 = G[pj2]
        g_3 = G[pj3]
        
        for j in range(m): 
            g1, g2, g3 = 0.0, 0.0, 0.0
            
            p1j = p1im[j]
            f1 = F[p1j]
            
            for k in range(j):
                p1k = p1im[k]
                if p1k == -1:
                    g1 += g_1*1.0
                    f1 = f_1*1.0
                else:
                    g1 += G[p1k]
            
            s1 += f1*exp(g1)
            
            p2j = p2im[j]
            f2 = F[p2j]
            
            for k in range(j):
                p2k = p2im[k]
                if p2k == -1:
                    g2 += g_1*1.0
                    f2 = f_1*1.0
                else:
                    g2 += G[p2k]
            
            s2 += f2*exp(g2)
            
            p3j = p3im[j]
            f3 = F[p3j]
            
            for k in range(j):
                p3k = p3im[k]
                if p3k == -1:
                    g3 += g_1*1.0
                    f3 = f_1*1.0
                else:
                    g3 += G[p3k]
                    
            s3 += F[p3j]*exp(g3)
            
        F1[i]=s1+s2+s3
        
    return F1/sum(F1)*sum(F)

\end{lstlisting}

\item\lstinline[basicstyle=\normalsize\ttfamily]{'parameters.py'}:
\begin{lstlisting}
from numpy import sign,log10

Om = 0.1 # Rabi frequency
Ng = 1e6 # Number of gain molecules
gK = 2 # K and K' decay rate
gnr = 5 # Nonradiative decay rate
G12 = 200 # Polarization relaxation rate

pr = 3 # Propagation rate
sr = 0.01 # Seeding rate due to spontaneous emission
rr = 200 # Envelope renormalization rate

t0 = 1.5 # Pump peak arrival time
tau = 0.25 # Pump peak duration
gmax = 100 # Peak pumping rate

dt = 5e-3 # Fixed time step for the Euler-Heun solver

noise = 0.01 # Additive noise in the K/K' mode amplitudes

a0 = 0.0+0.0j # Initial K mode amplitude
b0 = 0.0+0.0j # Initial K' mode amplitude

tspan = (0.0,3.5) # Time span

sv = 0 # Save solution: 0 - no, 1 - only amplitudes, 2 - everything

pm = 2 # Plot mode: 0 - none, 1 - basic, 2 - full

rep = 1 # Number of repetitions with the same parameters

show = 1 # Show plot (pause after each loop): 0 - no, 1 - yes

# If rep > 1 and show == 0, the program stops after 'rep' repetitions.
# Otherwise, it will keep running and reading updated parameter values.


def get_params(rx,ry,b,g):
    
    gp = max(g) # Amplification upon propagation in pumped areas
    lp = abs(min(g)) # Attenuation upon propagation outside pumped areas
    G=(sign(g)*1.0+1.0)/2.0 # Binary pump pattern
    F0 = G*0.0+1.0 # Initial envelope
    F0*=len(F0)/sum(F0)
    
    # Title of the plot
    pt = r'$\Omega$ = '+'%.3f'%Om + r'    $N_g$ = '+\
               '%.1f'%(10**(log10(Ng)-int(log10(Ng))))+\
               ' x 10'+r'$^'+str(int(log10(Ng)))+'$    '+\
               r'$\gamma_K = $'+'%.1f'%gK+\
               r'    $\gamma_{nr}$ = '+'%.1f'%gnr+\
               r'    $\Gamma_{12}$ = '+'%.0f'%G12+\
               r'    $t_0$ = '+'%.2f'%t0+\
               r'    $\tau$ = '+'%.2f'%tau+\
               r'    $g_{max}$ = '+'%.0f'%gmax+\
               r'    $\delta t$ = '+'%.4f'%dt+\
               r"    noise = "+'%.4f'%noise+'\n'\
               r'    propagation rate = '+'%.4f'%pr+\
               r'    seeding rate = '+'%.4f'%sr+\
               r'    envelope renormalization rate = '+'%.0f'%rr+\
               r'    amplification upon propagation = '+'%.4f'%gp+\
               r'    attenuation upon propagation = '+'%.4f'%lp+'\n'

    return Om,Ng,gK,gnr,G12,pr,sr,rr,t0,tau,\
           gmax,dt,noise,a0,b0,\
           tspan,F0,gp,lp,G,sv,pm,rep,show,pt

\end{lstlisting}

\item\lstinline[basicstyle=\normalsize\ttfamily]{'read_amplitudes.py'} (an additional code for statistical analysis):
\begin{lstlisting}
import matplotlib.pyplot as p
from numpy import absolute as ab
from numpy import log10 as lg
from numpy import angle as an
from numpy import array as a
from numpy import argmax as am
from numpy import arctan, pi, log10

nof = 100 # Number of files

disp_plot = 0 # Display plots: 0 - no, 1 - yes

#'''
fn_ = ['amplitudes_'+str(i)+'.txt' for i in range(nof)]

thav_,thpeak_,phav_,phpeak_=[],[],[],[]

for fn in fn_:
    file = open(fn,'r')
    lines = file.readlines()
    file.close()
    t,aK1,aK2=[],[],[]
    for line in lines:
        t+=[eval(line.split('\t')[0])]
        aK1+=[eval(line.split('\t')[1])+\
              1.0j*eval(line.split('\t')[2])]
        aK2+=[eval(line.split('\t')[3])+\
              1.0j*eval(line.split('\t')[4])]

    AK1 = ab(aK1)
    AK2 = ab(aK2)
    maxAK = a([max(AK1[i],AK2[i]) for i in range(len(t))])
    th = 2*arctan((AK1**2-AK2**2)/(AK1**2+AK2**2))/pi
    ph = (an(a(aK1))-an(a(aK2)))%(2*pi)/pi
    ipeak = am(maxAK)

    thpeak = th[ipeak]
    phpeak = ph[ipeak]

    thav = 2*arctan((sum(AK1**2)-sum(AK2**2))/sum(AK1**2+AK2**2))/pi
    phav = sum(ph*(AK1**2+AK2**2))/sum(AK1**2+AK2**2)

    thpeak_+=[thpeak]
    phpeak_+=[phpeak]
    thav_+=[thav]
    phav_+=[phav]

    if disp_plot == 1:
        p.figure(figsize=(16,6))
        p.subplot(131)
        p.title(r'$t_{peak}$ = '+str(t[ipeak])+' ps')
        p.plot(t,lg(AK1),'r',alpha=0.5)
        p.plot(t,lg(AK2),'b',alpha=0.5)
        p.ylabel('Amplitude')
        p.xlabel('Time (ps)')

        p.subplot(132)
        p.title(r'$\theta(t_{peak})$ = '+'%.4f'%(thpeak)+r'$\pi$,  '+\
                r'$\theta_{av}$ = '+'%.4f'%(thav)+r'$\pi$, ')
        p.scatter(t,th,c='k',s=0.5)
        p.ylabel(r'$\theta$ '+r'($\pi$)')
        p.xlabel('Time (ps)')
        
        p.subplot(133)
        p.title(r'$\phi(t_{peak})$ = '+'%.4f'%(phpeak)+r'$\pi$,  '+\
                r'$\phi_{av}$ = '+'%.4f'%(phav)+r'$\pi$, ')
        p.scatter(t,ph,c='k',s=0.5)
        p.xlabel('Time (ps)')
        p.ylabel('Relative phase '+r'($\pi$)')

        p.subplots_adjust(left=0.05,right=0.95)
        
        p.show()

#'''
# Uniform distribution example
'''
from numpy.random import random
thav_=random(nof)-0.5
thpeak_=random(nof)-0.5
phav_=random(nof)*2
phpeak_=random(nof)*2
#'''

fig=p.figure(figsize=(8,6))

from scipy.stats import kstest, uniform
from numpy import cumsum

p.suptitle('Histograms and KS test')

sp1=p.subplot(221)
cthav_,bthav_,i=sp1.hist(thav_,bins=len(thav_),range=(-0.5,0.5),alpha=0.25)
p.xlabel(r'$\theta_{av}$ '+r'($\pi$)')
p.xlim(-0.5,0.5)
p.ylabel('Occurrences')
csthav_=cumsum(cthav_)/sum(cthav_)
ax1=p.twinx(sp1)
x_=a(bthav_[:-1])+(bthav_[1]-bthav_[0])/2
ax1.plot(x_,csthav_,'k')
ax1.plot([-0.5,0.5],[0,1],'m')
xm = am(ab(csthav_-(x_+0.5)))
ax1.plot([x_[xm],x_[xm]],[x_[xm]+0.5,csthav_[xm]],'r')
p.xlim(-0.5,0.5)
p.ylim(-0.01,1.01)
test = kstest(a(thav_)+0.5, uniform.cdf)
p.title(r'$D_{max}$ = '+'%.4f'%(test.statistic)+\
           r', $p$-value = '+'%.4f'%(test.pvalue),fontsize=10)
p.ylabel('Cumulative sum')

sp2=p.subplot(222)
cphav_,bphav_,i=p.hist(phav_,bins=len(phav_),range=(0,2),alpha=0.25)
p.xlabel(r'$\phi_{av}$ '+r'($\pi$)')
p.xlim(0,2)
p.ylabel('Occurrences')
csphav_=cumsum(cphav_)/sum(cphav_)
ax2=p.twinx(sp2)
x_=a(bphav_[:-1])+(bphav_[1]-bphav_[0])/2
ax2.plot(x_,csphav_,'k')
ax2.plot([0,2],[0,1],'m')
xm = am(ab(csphav_-(x_*0.5)))
ax2.plot([x_[xm],x_[xm]],[x_[xm]*0.5,csphav_[xm]],'r')
p.xlim(0,2)
p.ylim(-0.01,1.01)
test = kstest(a(phav_)/2, uniform.cdf)
p.title(r'$D_{max}$ = '+'%.4f'%(test.statistic)+\
           r', $p$-value = '+\
        '%.4f'%(10**(log10(test.pvalue)-int(log10(test.pvalue))))+\
        ' x 10'+r'$^{'+str(int(log10(test.pvalue)))+r'}$',fontsize=10)
p.ylabel('Cumulative sum')

sp3=p.subplot(223)
cthpeak_,bthpeak_,i=p.hist(thpeak_,bins=len(thpeak_),range=(-0.5,0.5),alpha=0.25)
p.xlabel(r'$\theta_{peak}$ '+r'($\pi$)')
p.xlim(-0.5,0.5)
p.ylabel('Occurrences')
csthpeak_=cumsum(cthpeak_)/sum(cthpeak_)
ax3=p.twinx(sp3)
x_=a(bthpeak_[:-1])+(bthpeak_[1]-bthpeak_[0])/2
ax3.plot(x_,csthpeak_,'k')
ax3.plot([-0.5,0.5],[0,1],'m')
xm = am(ab(csthpeak_-(x_+0.5)))
ax3.plot([x_[xm],x_[xm]],[x_[xm]+0.5,csthpeak_[xm]],'r')
p.xlim(-0.5,0.5)
p.ylim(-0.01,1.01)
test = kstest(a(thpeak_)+0.5, uniform.cdf)
p.title(r'$D_{max}$ = '+'%.4f'%(test.statistic)+\
           r', $p$-value = '+'%.4f'%(test.pvalue),fontsize=10)
p.ylabel('Cumulative sum')

sp4=p.subplot(224)
cphpeak_,bphpeak_,i=p.hist(phpeak_,bins=len(phpeak_),range=(0,2),alpha=0.25)
p.xlabel(r'$\phi_{peak}$ '+r'($\pi$)')
p.xlim(0,2)
p.ylabel('Occurrences')
csphpeak_=cumsum(cphpeak_)/sum(cphpeak_)
ax4=p.twinx(sp4)
x_=a(bphpeak_[:-1])+(bphpeak_[1]-bphpeak_[0])/2
ax4.plot(x_,csphpeak_,'k')
ax4.plot([0,2],[0,1],'m')
xm = am(ab(csphpeak_-(x_*0.5)))
ax4.plot([x_[xm],x_[xm]],[x_[xm]*0.5,csphpeak_[xm]],'r')
p.xlim(0,2)
p.ylim(-0.01,1.01)
test = kstest(a(phpeak_)/2, uniform.cdf)
p.title(r'$D_{max}$ = '+'%.4f'%(test.statistic)+\
           r', $p$-value = '+\
        '%.4f'%(10**(log10(test.pvalue)-int(log10(test.pvalue))))+\
        ' x 10'+r'$^{'+str(int(log10(test.pvalue)))+r'}$',fontsize=10)
p.ylabel('Cumulative sum')

p.subplots_adjust(hspace=0.4,wspace=0.4)
fig.savefig('KS_test.jpg',format='jpg',dpi=600)
p.show()

    
\end{lstlisting}
\end{enumerate}

\bibliography{bibliography}